# Leveraging online shopping behaviors as a proxy for personal lifestyle choices: New insights into chronic disease prevention literacy

**Yongzhen Wang[1], Xiaozhong Liu[2], Katy Börner[3], Jun Lin[4], Yingnan Ju[3], Changlong Sun[4], and Luo Si[4]**

[1]Institute of Science of Science and S&T Management, Dalian University of Technology, Dalian, China

[2]Department of Computer Science, Worcester Polytechnic Institute, Worcester, MA, USA

[3]School of Informatics, Computing and Engineering, Indiana University, Bloomington, IN, USA

[4]Alibaba DAMO Academy, Hangzhou, China

* To whom correspondence may be addressed.

yongzhenwang@dlut.edu.cn (YW) and xliu14@wpi.edu (XL)

# Abstract

**Objective:** Ubiquitous internet access is reshaping the way we live, but it is accompanied by unprecedented challenges in preventing chronic diseases that are usually planted by long exposure to unhealthy lifestyles. This paper proposes leveraging online shopping behaviors as a proxy for personal lifestyle choices to improve chronic disease prevention literacy, targeted for times when e-commerce user experience has been assimilated into most people's everyday lives.

**Methods:** Longitudinal query logs and purchase records from 15 million online shoppers were accessed, constructing a broad spectrum of lifestyle features covering various product categories and buyer personas. Using the lifestyle-related information preceding online shoppers' first purchases of specific prescription drugs, we could determine associations between their past lifestyle choices and whether they suffered from a particular chronic disease.

**Results:** Novel lifestyle risk factors were discovered in two exemplars—depression and type 2 diabetes, most of which showed reasonable consistency with existing healthcare knowledge. Further, such empirical findings could be adopted to locate online shoppers at higher risk of these chronic diseases with decent accuracy [i.e., (area under the receiver operating characteristic curve) AUC=0.68 for depression and AUC=0.70 for type 2 diabetes], closely matching the performance of screening surveys benchmarked against medical diagnosis.

**Conclusions:** Mining online shopping behaviors can point medical experts to a series of lifestyle issues associated with chronic diseases that are less explored to date. Hopefully, unobtrusive chronic disease surveillance via e-commerce sites can grant consenting individuals a privilege to be connected more readily with the medical profession and sophistication.

# Keywords

# Introduction

Chronic diseases, such as heart disease, cancer, and diabetes, are the leading causes of mortality and disability across the globe. For instance, in the United States, 6-in-10 adults live with at least one chronic disease, and 70% of deaths and nearly 75% of aggregate healthcare spending are due to chronic diseases[1]. Similarly, in China, the largest developing country with a population of 1.4 billion, chronic diseases account for an estimated 80% of deaths and 70% of disability-adjusted life years lost[2]. Although early detection of chronic diseases can trigger timely cures and prolonged survival, regular health checkups can be cost-prohibitive[3] and incentive-deficient[4]. To alleviate underdiagnosis and undertreatment, healthy lifestyles such as a balanced diet[5], smoking cessation[6], physical exercise[7], and alcohol withdrawal[8] have long been advocated as preventive measures against chronic diseases. Still, new avenues to granular, actionable data profiling for lifestyles in the goal of chronic disease prevention are needed, since booming digital platforms continuously revolutionize our everyday lives through the internet of everything[9].

Recently, it was reported that predicting the early risk of chronic kidney disease in patients with diabetes via real-world data showed enhanced performance vs. clinical data[10]. In addition, similar findings were also drawn from a series of studies built upon social media sites, on which social media users' lifestyle changes could be documented unobtrusively. For example, Twitter corpora were used to develop a statistical classifier to provide a risk estimate of how depressed a user was[11]. After that, Reddit posts were used to create models capable of automatically detecting anxiety disorders[12]. Over the same period, Instagram photos were used to establish useful psychological indicators to reveal predictive markers of depression[13]. Later, Facebook languages were used to predict occurrences of depression-related events in users' medical records[14]. Unlike dedicated and strictly controlled medical research, these data-driven studies did not bring out any hypothetical questions about chronic diseases, while stimulating new opportunities to achieve lifestyle-oriented data profiling based on digital platforms.

This paper proposes leveraging online shopping behaviors as a proxy for personal lifestyle choices, aiming to explore unrevealed lifestyle risk factors and accordingly open an innovative way to chronic disease risk prediction. The reasons and motivations are as follows. Online shopping has had a profound impact on the ways people live their lives—the benefits of online shopping are becoming seemingly endless and has changed the culture and behaviors of shoppers everywhere[15]. Amid the flourishing era of e-commerce, the number of online shoppers worldwide has now reached 2.14 billion[16], and the global online retail sales are expected to increase up to 5.4 trillion dollars by 2022[17]. So here comes the question: is there a connection between online shoppers and healthcare consumption? Yes, a Harris Interactive study demonstrates that well online healthcare consumers make up approximately 60% of the consumers searching for health information online—they search for preventive medicine and wellness information in the same way they look for news, stock quotes, and products; for those newly diagnosed, they will search frenetically in the first few weeks following their diagnosis, and many of them will cast a wide net for medical help online[18]. It is conceivable that the incorporation of online shopping behaviors into chronic diseases prevention will grow in practice and importance as more people go shopping online.

This paper selects China, the world's top retail market with 35.3% of sales taking place online[19], as the testbed. Meanwhile, Alibaba, the bellwether of China's e-market with a 53.3% share and 758 million active users[20], is chosen as the data source. Alibaba situates a great wealth of goods and services in the context of almost every aspect of Chinese living consumption, ranging from explicit features, such as food intake and entertainment spending, to implicit ones, such as body size and clothing preference. On this basis, this paper sheds spotlights on two representative chronic diseases—depression and type 2 diabetes that affect and can be affected by personal lifestyle choices[21, 22]—as two case studies. The reasons and motivations are as follows. In China, rapid social and economic changes quickened the pace of living and caused a general increase in psychological pressure and stress. Recently, the prevalence rate of depressive symptoms among Chinese adults was estimated to be 37.9%[23], but the recognition rate of depression in Shanghai, a Tier-1 city of China, was only 21%, far below the world average[24]. Similarly, along with China's urbanization and rising living standards during the past decades, the prevalence rate of type 2 diabetes soared from less than 1% in the 1980s to dramatically more than 10%[25]. Unfortunately, more than 60% of Chinese adults with type 2 diabetes were unaware of their diagnosis[26].

This paper addresses the following two research questions:

- Can online shopping behaviors reveal lifestyle risk factors associated with chronic diseases?
- Can online shopping behaviors be utilized to accomplish chronic disease risk prediction?

To the best of our knowledge, we have made the first attempt to translate online shopping behaviors to lifestyle-oriented data profiling in order to improve chronic disease prevention literacy. As one of the deliverables of this paper, we find that *"if female online shoppers aged 15-to-24 used to (i) shop frequently, (ii) endure financial pressure, and (iii) spend more on healthcare products, alcoholic drinks, haircare services, and reading materials, but (iv) pay less for phone expenses and women's clothing, they could face a higher chance of developing depression"*. Comparably, such an empirical finding could be quite intractable to obtain for other digital platforms, even though some of them, such as Twitter, Reddit, Instagram, and Facebook, proved to identify online users with psychological abnormality via user-generated content, whose original intention was for social sharing[11-14].

## Materials and Methods

*Overview*

This paper surveyed 15 million Alibaba users at random within a year span from January 1 to December 31, 2018; however, users were excluded if they had bought any prescription drug prior to 2018 to guarantee an unbiased analysis for the later onset of chronic diseases. The yearlong span was partitioned into an eight-month observation period (i.e., January through August) and a four-month performance period (i.e., September through December). Notably, only information generated within the observation period, including 6 billion query logs and 3.2 billion purchase records, were used as input for lifestyle-oriented data profiling. As for the two case studies, users who bought prescription drugs for depression and type 2 diabetes (shown in Figure 1, listed in Supplementary Appendix A) during the performance period for the first time were defined as depressed and diabetic users, respectively. In contrast, those who did not make such purchases over the performance period were specified as control users. The sampling ratios of users with respect to depressed vs. control and diabetic vs. control were balanced at 1-to-19 and 1-to-9, respectively. These settings allowed us to compare different online shoppers' personal lifestyle choices across the same time window and simulate both prevalence rates of depression and type 2 diabetes in China (i.e., approximately 5% and 10%)[27, 28].

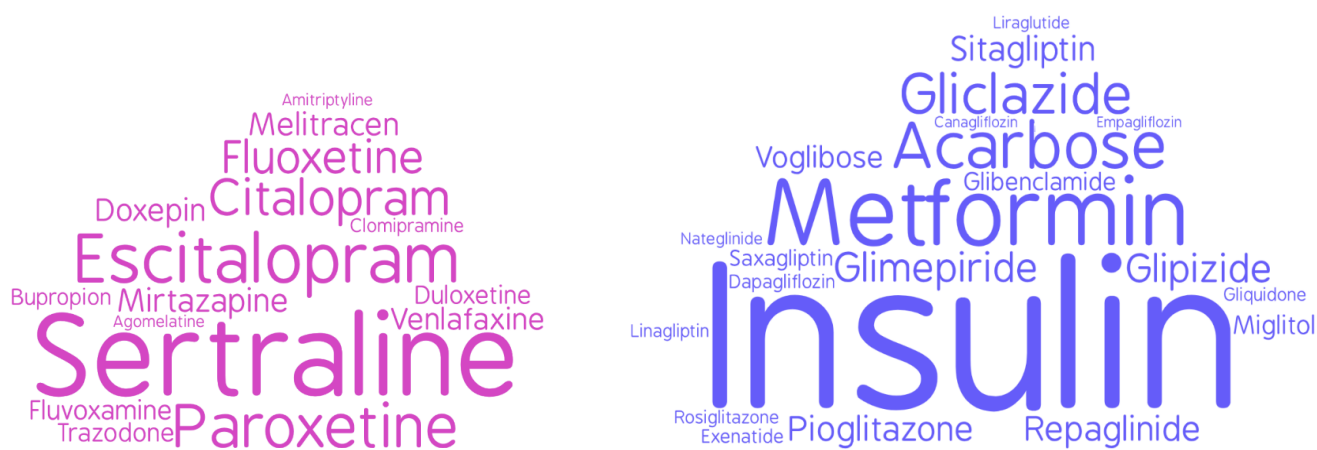

**Figure 1.** Prescription drugs found in Alibaba users' purchase records that are relevant to depression (*Left*) and type 2 diabetes (*Right*). Here, the generic name of each prescription drug is listed for simplicity. Note that larger fonts indicate higher relative purchase frequencies.

*Data Collection*

From 15 million Alibaba user accounts, we retrieved demographics (sex and age) and a total of 6 billion query logs and 3.2 billion purchase records beginning on January 1, 2018 and ending on December 31, 2018. For the two case studies, 3071 depressed and 3936 diabetic users were identified. The user characteristics are reported in Table 1. Note that the control users (also the majority of the subjects of this research) reveal some basic characteristics of Chinese online shoppers. Overall, depressed and diabetic users are older (Student's t-test: for depression,

$t'$=14.874, $df$=4431.893, $p$<0.001; for type 2 diabetes, $t'$=34.899, $df$=5687.344, $p$<0.001) and include a smaller proportion of women (Pearson's Chi-squared test: for depression, $\chi^2$=333.180, $df$=1, $N$=13071, $p$<0.001; for type 2 diabetes, $\chi^2$=107.035, $df$=1, $N$=13936, $p$<0.001). Furthermore, they both tend to place more orders (Mann-Whitney U test: for depression, $U$=14661834, $z$=-3.790, $p$<0.001; for type 2 diabetes, $U$=17241491, $z$=-11.406, $p$<0.001), while diabetic users tend to search for products less frequently (Mann-Whitney U test: for depression, $U$=15013757.5, $z$=-0.283, $p$=0.777>0.05; for type 2 diabetes, $U$=18322854.5, $z$=-4.472, $p$<0.001). In order to model the two case studies in a real-world scenario, for each depressed user, we randomly selected another 19 control users of the same demographics, yielding a sample of 3071+19×3071=61420 users (i.e., 1-to-19 for depressed vs. control) for investigating depression[27]. Analogously, a sample of 3936+9×3936=39360 users (i.e., 1-to-9 for diabetic vs. control) was chosen for investigating type 2 diabetes[28]. Each sample was then divided into several parts for subgroup analysis, keeping depressed/diabetic users and their control counterparts within the same subsample, given the fact that online shopping behaviors differed significantly by sex and age (explained in Supplementary Appendix B).

**Table 1.** Online shopper characteristics. Here, 10000 control (Alibaba) users were randomly sampled for ease of comparison. Differences in age were tested by the Student's t-test with equal variances not assumed, percent female by the Pearson's Chi-squared test, query count and purchase count by the Mann-Whitney U test, in comparison with the control users.

| **Descriptive** | **Depressed** | **Diabetic** | **Control** |
| --- | --- | --- | --- |
| #Subject | 3071 | 3936 | 10000 |
| Age ($M \pm SD$) | 34.4±10.5* | 38.5±11.8* | 31.3±8.7 |
| Female (%) | 37.9* | 47.0* | 56.7 |
| Monthly #Query ($M \pm SD$) | 48.9±58.5 | 44.5±50.4* | 46.6±50.3 |
| Monthly #Purchase ($M \pm SD$) | 13.8±11.8* | 15.9±13.5* | 13.1±12.2 |

*Significant difference ($\alpha$=0.05)

$M$: mean; $SD$: standard deviation

## Feature Engineering and Selection

Two types of lifestyle features were engineered. One type was to unfold online shoppers' daily purchases explicitly (e.g., food intake and entertainment spending), while another type was to characterize their living consumption implicitly (e.g., body size and clothing preference). More concretely, 135 explicit lifestyle features were generated by pooling 3.2 billion purchase records into a list of sales statistics according to Alibaba product categorization. Meanwhile, 115 implicit lifestyle features were constructed by creating a variety of buyer personas from 6 billion query logs and 3.2 billion purchase records using Alibaba off-the-shelf data mining technologies. Notably, buyer personas for Alibaba users under 18 years old (≈0.5% of the subjects of this research) were not generated due to specific data policy. All these features were discretized to group data into different bins, and we prudently eliminated the collinearity among them. As for the two case studies, only the lifestyle features (listed in Supplementary Appendix C) showing a correlation with depression/type 2 diabetes in the Chi-squared test of independence (quantified by a humble significance threshold $p<0.1$) were retained for further analysis.

## Regression Analysis

Consider a regression model with multiple explanatory variables $x_1, x_2, \cdots, x_m$ and one binary explained variable $y$, aiming to estimate the probability of online shoppers suffering from a particular chronic disease $\pi=\Pr(y=1|x_1, x_2, \cdots, x_m)$. For clarity, $x_1, x_2, \cdots, x_m$ represents an array of $m$ lifestyle features of an online shopper, and $y=1$ indicates that he/she will be diagnosed with the focal chronic disease in the future (otherwise $y=0$). Without loss of generality, a linear relationship is assumed between $x_1, x_2, \cdots, x_m$ and the log odds of $y=1$, and multiple logistic regression[29] can be established as:

$$\ln\left(\frac{\pi}{1-\pi}\right) = \beta_0 + \beta_1 x_1 + \beta_2 x_2 + \cdots + \beta_m x_m$$

Here, odds ratio (*OR*) is used to interpret the standardized regression coefficients $\beta_1, \beta_2, \cdots, \beta_m$. For example, the *OR* for a one-unit increase in $x_1$ is $\exp(\beta_1)$—there is a $[\exp(\beta_1)-1]\times 100\%$ increase or decrease in the odds of $y=1$ when $x_1$ increases by one unit. If $\exp(\beta_1)>1$, then $x_1$ is positively [or negatively if $\exp(\beta_1)<1$] associated with the odds of $y=1$.

*Statistical Power Analysis*

A priori power analysis was performed for subsample size estimation. The effect size in this paper was expected to be *OR*=1.49 (or inverted 0.67), considered to be small using Cohen's criteria[30]. With α=0.05, power=0.8, probability of null hypothesis being true=0.5, proportion of variance for other covariates=0.8, the projected subsample size needed with this effect size was approximately 1068 for logistic regression. Therefore, we excluded the depression subsamples aged above 55 and type 2 diabetes subsamples aged above 65 due to insufficient statistical power, as shown in Table 2. Consequently, only 59140 (≈96.3%) and 38420 (≈97.6%) Alibaba users were retained for the two case studies of depression and type 2 diabetes, respectively.

**Table 2.** Subsample statistics. Note that the subsamples with a strikethrough are removed due to insufficient statistical power.

| Chronic Disease | Sex | Age | | | | | | Total |
|---|---|---|---|---|---|---|---|---|
| | | 15-24 | 25-34 | 35-44 | 45-54 | 55-64 | 65-74 | |
| Depression | Female | 5660 | 6500 | 6200 | 3900 | ~~920~~ | ~~80~~ | 59140 |
| | Male | 7400 | 13840 | 10300 | 5340 | ~~920~~ | ~~360~~ | |
| Type 2 Diabetes | Female | 2960 | 6400 | 4940 | 2870 | 1160 | ~~160~~ | 38420 |
| | Male | 1600 | 5520 | 5930 | 5040 | 2000 | ~~780~~ | |

Remarkably, for type 2 diabetes, the subsamples aged above 55 make up 10.42% of the total, quite close to the share of online shoppers over 50 years old in China (i.e., 10.7%)[31]. But for depression, the subsamples aged above 55 account for only 3.71% of the total. In addition, the age distribution of online shoppers varies significantly from country to country. Take the United States as an example, 29% of online shoppers are 55 and older[32], almost three times the number in China.

*Multiple Comparisons*

In this paper, a lifestyle feature was considered a lifestyle risk factor if its estimated regression coefficient was significantly different from zero, i.e., if it was significantly associated with chronic disease onset. The Benjamini-Hochberg adjusted p-value[33] was then applied to control false-positive errors among the discovered lifestyle risk factors at a level below 5%.

*Prediction Procedure*

We employed a support vector machine[34] with a misclassification cost parameter (illustrated in Supplementary Appendix D) to develop predictive classifiers to provide risk estimates for chronic diseases. All selected lifestyle features were used as predictors. In practice, we deliberately chose AUC (area under the receiver operating characteristic curve), a widely used indicator suitable for describing the classification accuracy over imbalanced classes, as our evaluation metric[35]. Moreover, 10-fold cross-validation was adopted to avoid over-fitting as follows. Each subsample was partitioned into ten stratified folds—one predictive classifier was trained using nine folds and was evaluated using the remaining held-out fold. This process was repeated ten times, each time with a different held-out fold, and then the results were averaged, yielding cross-validated out-of-sample AUC for performance assessment.

## Exploration of Lifestyle Risk Factors

To reveal associations between online shoppers' past lifestyle choices and whether they suffered from a particular chronic disease or not, we first mapped their historical query logs and purchase records into a wide array of interpretable and fine-grained lifestyle features, then chose multiple logistic regression[29] as the analytical process allowing for an explanation of this data-driven exploration (detailed in **Materials and Methods,** *Regression Analysis*). Here, the lifestyle features significantly associated with chronic disease onset can be regarded as lifestyle risk factors. Figure 2 demonstrates all lifestyle risk factors derived from the two case studies of depression and type 2 diabetes when demographics are controlled for. More information about the regression output can be found in Supplementary Appendix E.

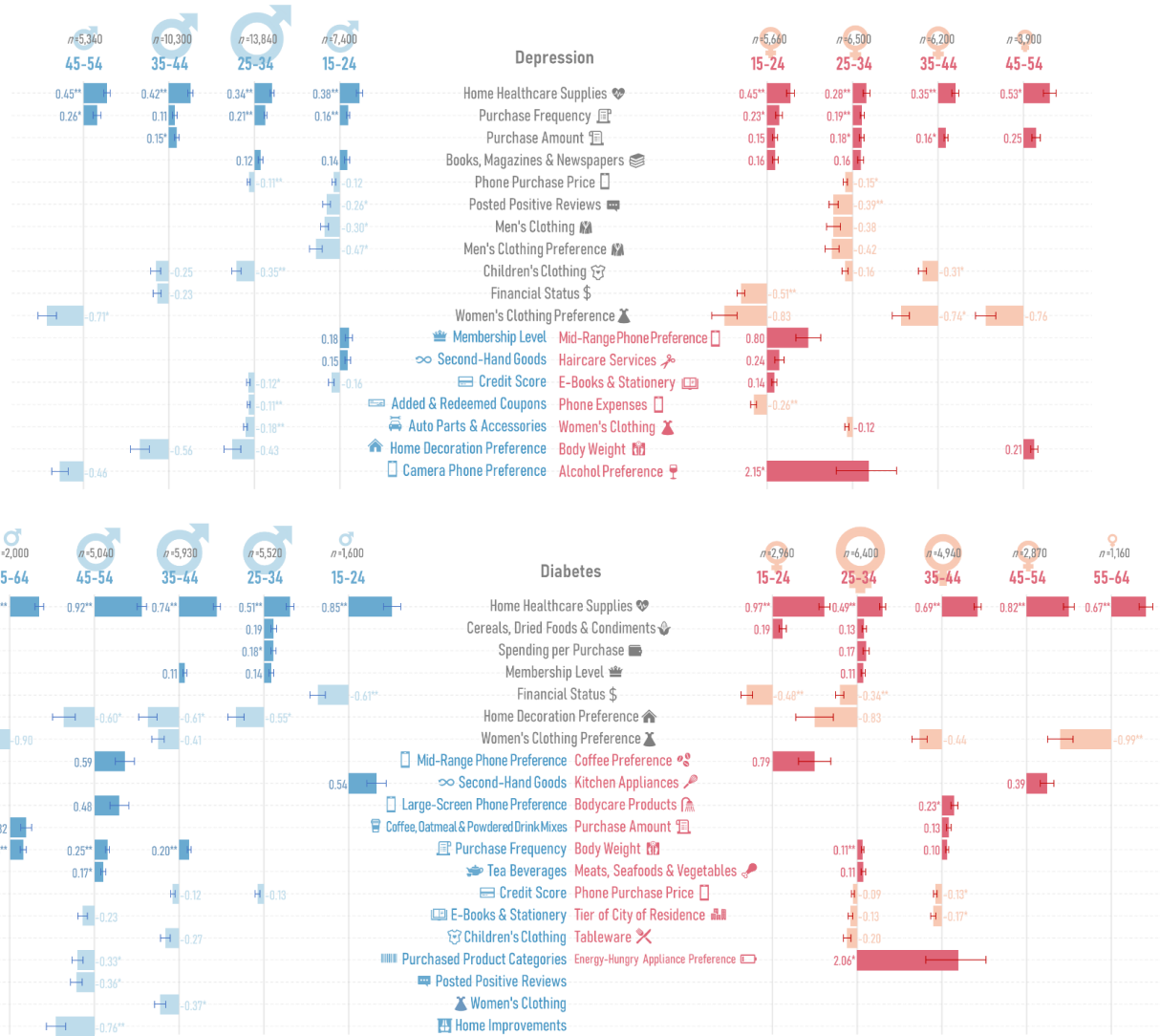

**Figure 2.** Discovered lifestyle risk factors associated with depression (*Top*, *N*=59140) and type 2 diabetes (*Bottom*, *N*=38420) when demographics are controlled for, reported as standardized regression coefficients (bars) and standard errors (two-end lines). Note that each column represents a subgroup analysis using multiple logistic regression, with the related sex, age, and size *n* of each subsample marked on top. Multiple significance tests are conducted by the Benjamini-Hochberg (BH) procedure.

*$p^{BH}$<0.01, **$p^{BH}$<0.001, otherwise $p^{BH}$<0.05

*Case 1: Depression*

For depression, the distribution of discovered lifestyle risk factors presents a conspicuous downtrend with the aging of online shoppers, roughly aligning with the distribution of ages at onset of the first major depressive episode[36]. Understandably, the shopping category most closely associated with depression falls on healthcare products such as over-the-counter drugs and contraceptives [e.g., focal lifestyle feature@$_\text{age}^\text{sex}$ *home healthcare supplies*@$_{15-24}^\text{female}$, odds ratio *OR*=1.570, standardized regression coefficient ± standard error β=0.451±0.079, Benjamini-Hochberg (BH) adjusted p-value $p^\text{BH}$<0.001].

Overall, depressed online shoppers from 15-to-54 years old manifest a stronger user viscosity to e-commerce sites than otherwise healthy counterparts (e.g., *purchase frequency*@$_{15-24}^\text{male}$, *OR*=1.173, β=0.159±0.030, $p^\text{BH}$<0.001; *purchase amount*@$_{25-34}^\text{female}$, *OR*=1.195, β=0.178±0.051, $p^\text{BH}$=0.004). Reasonably, e-commerce platforms can provide a superior shopping venue for individuals with depression potentials who are less proactive to participate in face-to-face social interactions[37]. Further, 15-to-24 year-old depressed online shoppers are inclined to show special interest in reading materials (e.g., *books, magazines & newspapers*@$_{15-24}^\text{female}$, *OR*=1.171, β=0.158±0.055, $p^\text{BH}$=0.021). Another impressive observation lies in that female depressed online shoppers aged 25-to-34 tend to buy clothing for the opposite sex and children less frequently (e.g., *men's clothing*@$_{25-34}^\text{female}$, *OR*=0.681, β=-0.384±0.141, $p^\text{BH}$=0.038; *children's clothing*@$_{25-34}^\text{female}$, *OR*=0.856, β=-0.155±0.060, $p^\text{BH}$=0.043), while male depressed online shoppers aged 25-to-44 are less prone to pay their attention to kids' clothes and home improvements (e.g., *children's clothing*@$_{25-34}^\text{male}$, *OR*=0.704, β=-0.351±0.090, $p^\text{BH}$<0.001; *home decoration preference*@$_{25-34}^\text{male}$, *OR*=0.649, β=-0.432±0.163, $p^\text{BH}$=0.048). To a significant extent, these depressed online shoppers are more likely to be single, moderately supporting the traditional perspective that marriage or cohabiting can promote psychological well-being[38].

There are some other findings consistent with existing literature on depression. For example, a glaring positive association stands between depression and alcohol consumption among female online shoppers of 15-to-24 years old (*alcohol preference*@$_{15-24}^\text{female}$, *OR*=8.612, β=2.153±0.598, $p^{BH}$=0.005), reiterating the causal inference that increased alcohol involvement raises the incidence rate of depression for adolescents and youth[39]. Moreover, female depressed online shoppers aged 15-to-24 are more likely to suffer from hair problems (*haircare services*@$_{15-24}^{\text{female}}$, *OR*=1.268, β=0.237±0.089, $p^{BH}$=0.032), confirming the popular perception that hair damage and mental disorders often occur in combination[40]. As for female depressed online shoppers at the ages of 45-to-54, they can be exposed to greater risk of being overweight (*body weight*@$_{45-54}^{\text{female}}$, *OR*=1.239, β=0.214±0.071, $p^{BH}$=0.026), matching the bi-directional relationship between depression and obesity in middle-aged women[41]. In addition, several discovered lifestyle risk factors relevant to personal finances (e.g., *financial status*@$_{15-24}^{\text{female}}$, *OR*=0.602, β=-0.507±0.074, $p^{BH}$<0.001; *credit score*@$_{15-24}^{\text{male}}$, *OR*=0.849, β=-0.164±0.053, $p^{BH}$=0.015) conform to the subjective intuition that lacking money usually leads to magnified feelings of anxiety and depression in many people[42].

*Case 2: Type 2 Diabetes*

For type 2 diabetes, the number of discovered lifestyle risk factors for female online shoppers and their male counterparts peaks at the ages of 25-to-34 and 45-to-54, respectively. This discrepancy partly echoes the sex difference that women have a higher prevalence of type-2 diabetes in youth while men see a higher prevalence in midlife[43].

Just like in the case of depression, the most salient shopping category associated with type 2 diabetes points to healthcare products (e.g., *home healthcare supplies*@$_{15-24}^{\text{female}}$, *OR*=2.639, β=0.970±0.105, $p^{BH}$<0.001). Also, a majority of diabetic online shoppers display high consumer inertia with respect to e-commerce sites (e.g., *membership level*@$_{25-34}^{\text{female}}$, *OR*=1.118, β=0.112±0.042, $p^{BH}$=0.043; *purchase frequency*@$_{35-44}^{\text{male}}$, *OR*=1.219, β=0.198±0.037, $p^{BH}$<0.001). On the whole, however, diabetic online shoppers favor much more diet-related purchasing activities (e.g., *cereals, dried foods & condiments*@$_{15-24}^{\text{female}}$, *OR*=1.205, β=0.187±0.065, $p^{BH}$=0.047; *coffee, oatmeal & powdered drink mixes*@$_{55-64}^{\text{male}}$, *OR*=1.377, β=0.320±0.106, $p^{BH}$=0.012). Clearly, food intake patterns play a crucial role during the progression toward type 2 diabetes[44]. In addition, female diabetic online shoppers from 25-to-34 and 45-to-54 years old exhibit a unique preference for household appliances (*energy-hungry appliance preference*$@_{25-34}^{\text{female}}$, *OR*=7.862, β=2.062±0.663, $p^{\text{BH}}$=0.003; *kitchen appliances*$@_{45-54}^{\text{female}}$, *OR*=1.484, β=0.395±0.127, $p^{\text{BH}}$=0.031).

There exist some other findings congruent with previous studies on type 2 diabetes. For example, 15-to-34 year-old diabetic online shoppers tend to endure financial pressure (e.g., *financial status*$@_{15-24}^{\text{female}}$, *OR*=0.617, β=-0.483±0.106, $p^{\text{BH}}$<0.001; *credit score*$@_{25-34}^{\text{male}}$, *OR*=0.881, β=-0.127±0.045, $p^{\text{BH}}$=0.039), corroborating the social inequality that poverty enlarges the likelihood of developing type 2 diabetes[45]. Another plausible conclusion can be that a larger proportion of female diabetic online shoppers at the ages of 25-to-44 may settle in big cities (e.g., *tier of city of residence*$@_{35-44}^{\text{female}}$, *OR*=0.841, β=-0.173±0.051, $p^{\text{BH}}$=0.009), aligning with the sharp increase in China's type 2 diabetes prevalence rate attached to its rapid diffusion of urbanization[46]. Moreover, female diabetic online shoppers aged 25-to-44 can be peculiarly susceptible to obesity (e.g., *body weight*$@_{25-34}^{\text{female}}$, *OR*=1.111, β=0.105±0.023, $p^{\text{BH}}$<0.001), corresponding to the fact that excessive weight gain is a typical expression emanating from the progression of type 2 diabetes[47]. As for male diabetic online shoppers aged 35-to-44, fewer purchasing activities are observed when it comes to women's and kids' wear and home decoration (e.g., *women's clothing*$@_{35-44}^{\text{male}}$, *OR*=0.688, β=-0.374±0.107, $p^{\text{BH}}$=0.004; *children's clothing*$@_{35-44}^{\text{male}}$, *OR*=0.764, β=-0.269±0.098, $p^{\text{BH}}$=0.029; *home decoration preference*$@_{35-44}^{\text{male}}$, *OR*=0.542, β=-0.612±0.176, $p^{\text{BH}}$=0.002), suggesting that they are more likely to remain single. This outcome adds empirical evidence to the recent research revealing that marriage can, in a way, protect middle-aged and older men from adult-onset diabetes[48].

*Executive Summary*

This section presents a data-driven workflow to explore lifestyle risk factors underlying online shopping behaviors, with the purpose of improving chronic disease prevention literacy to cope with the ongoing digitalization of daily life. The discovered lifestyle risk factors involve a variety of product categories and buyer personas, most of which demonstrate reasonable consistency with existing healthcare knowledge. Also, these empirical findings can, to a certain degree, allow medical experts to capture the dynamics of chronic disease severity across time with a richness that is unavailable to conventional health checkups delivered at discrete points of time. More importantly, mining online shopping behaviors can point medical experts to a series of lifestyle issues associated with chronic diseases that are less explored to date.

## Chronic Disease Risk Prediction

To identify online shoppers at higher risk of chronic diseases, we built a support vector machine (SVM)[34] as the predictive classifier based on their past lifestyle choices that were reflected in their historical query logs and purchase records, along with the use of 10-fold cross-validation to avoid over-fitting (detailed in **Materials and Methods**, *Prediction Procedure*). This model employed the interpretable and fine-grained lifestyle features as predictors, and was evaluated by comparing the estimated probability of online shoppers suffering from a particular chronic disease against the actual presence or absence of related prescription drugs in their purchase records. By varying the threshold of predicted probabilities for classification, a receiver operating characteristic (ROC) curve was uniquely determined. The area under the ROC curve (AUC) was calculated as a proxy for the accuracy of the early risk prediction of chronic diseases[35].

*Performance Analysis*

Figure 3 illustrates the prediction performance for the two case studies of depression and type 2 diabetes when demographics are controlled for. Notably, all predictors here are divided into two disjoint subsets to validate the discovered lifestyle risk factors empirically. On average, these lifestyle risk factors can result in cross-validated out-of-sample AUC of 0.678 and 0.695 when applied to depression and type 2 diabetes risk prediction, respectively, falling just short of the customary threshold for good discrimination (i.e., 0.7). Meanwhile, their predictive power significantly outperforms that of the placebos (Wilcoxon signed-rank test: for depression, $z$=-2.521, $p$=0.012; for type 2 diabetes, $z$=-2.803, $p$=0.005), and combining both subsets cannot improve final predictive accuracy by a substantial margin (Wilcoxon signed-rank test: for depression, $z$=-1.260, $p$=0.208>0.05; for type 2 diabetes, $z$=-1.174, $p$=0.241>0.05).

To sum up, the discovered lifestyle risk factors can capture most of the depression-related or type 2 diabetes-related variance rooted in online shopping behaviors. Consequently, the placebos will be eliminated from all predictors to reduce model complexity. More results about the prediction procedure can refer to Supplementary Appendix F.

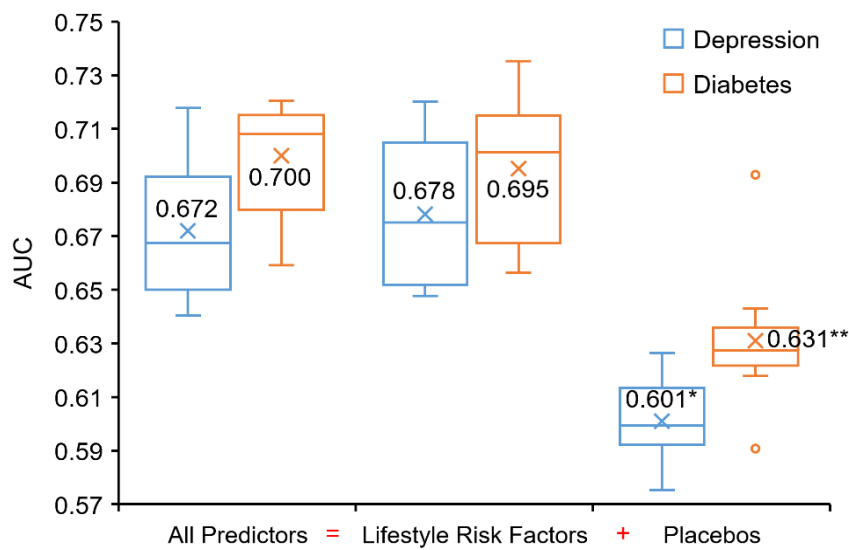

**Figure 3.** Online shopping behaviors-based prediction performance (via SVM) in early risk of depression ($N$=8) and type 2 diabetes ($N$=10) when demographics are controlled for. Statistical analysis is conducted by the Wilcoxon signed-rank test, in comparison with the lifestyle risk factors.

*$p$<0.05, **$p$<0.01

AUC: area under the receiver operating characteristic curve

## Comparison with Well-Established Screening Surveys

Figure 4 compares our proposed online shopping behaviors-based predictive classifiers against existing screening surveys for depression and type 2 diabetes, including an electronic medical records (EMRs)-based detection method of depression[49] and several diabetes risk assessment instruments[50] (detailed in Supplementary Appendix G). These baselines all select medical diagnosis as the gold standard for benchmarking. For depression, our proposed predictive classifiers perform nearly as well as those based on "diagnostic code", "problem list", and "medication list" jointly (i.e., three fields on EMRs) when a low false-positive rate is required. Even when it comes to a more relaxed restriction on false-positive errors, they still match closely with the baseline resorting solely to "problem list". Notably, the EMRs collected by Trinh et al.[49] originate from primary care patients, whereas the online shopping behaviors of this study come from a general population. As for type 2 diabetes, our proposed predictive classifiers display an obvious advantage over four Western-oriented risk prediction models (for American, Danish, Dutch, and Finnish) regarding relatively strict classification thresholds (i.e., a high cutoff probability for classifying a subject as positive) and, meanwhile, yield performance similar to three Asian-oriented ones (for Chinese, Indian, and Thai) regarding fairly lax classification criteria. Moreover, each of the seven baselines tested by Gao et al.[50] contains some strong predictors such as the family history of diabetes[51] and known condition of hypertension[52]. On the contrary, no hypothetical guidance has been applied to the feature engineering and selection of this research. Therefore, it is convincing to conclude that our proposed online shopping behaviors-based predictive classifiers can provide risk estimates for depression and type 2 diabetes as accurately as screening surveys benchmarked against medical diagnosis.

*Executive Summary*

This section elaborates how to pre-screen online shoppers for the prevention of chronic diseases by leveraging their "lifestyle profiles" documented in their historical query logs and purchase records. The growing population of online shoppers can thus be empowered to access low-cost early risk prediction of chronic diseases and accordingly jump-start personalized health intervention at the proper time—for example, e-commerce platforms should lessen or stop advertising sugary cereals and sweetened drinks to online shoppers who have already been informed of a high chance of heart disease. Hopefully, unobtrusive chronic disease surveillance via e-commerce sites is expected to be available for consenting individuals to be connected more readily with essential medical resources, cooperating with professional treatments and nursing to attain more guaranteed wellness.

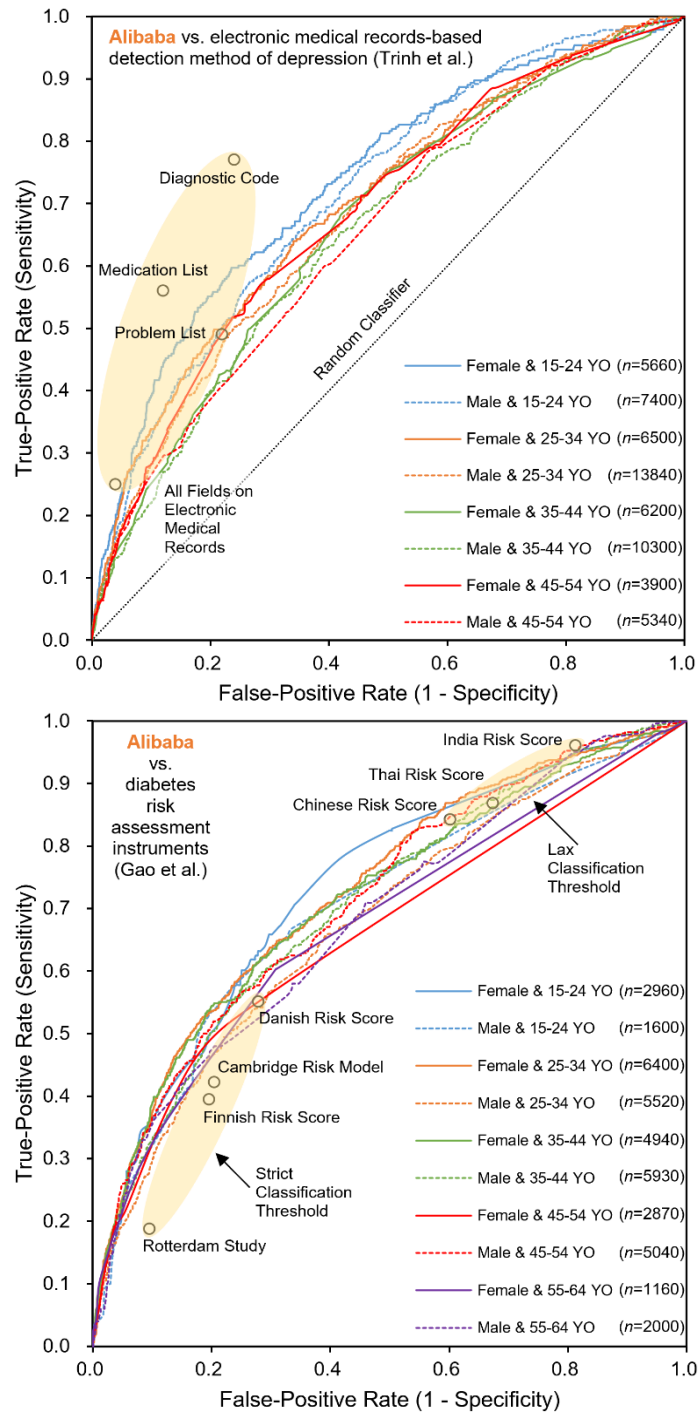

**Figure 4.** In-sample ROC curves of online shopping behaviors-based predictive classifiers (via SVM) for depression (*Left*, *N*=59140) and type 2 diabetes (*Right*, *N*=38420) when demographics are controlled for. The points as combinations of true- and false-positive rates (shaded areas) are reported by the previous screening surveys—an electronic medical records-based detection method of depression (*N*=427) and several diabetes risk assessment instruments (*N*=4336).

ROC: receiver operating characteristic, YO: years old

# Conclusions and Outlook

*Significance Statement*

Digitalization of daily life calls for new insights into chronic disease prevention literacy. This paper shows that online shopping behaviors can be leveraged as a proxy for personal lifestyle choices to discover novel lifestyle risk factors and provide accurate risk estimates for chronic diseases. It details relevant data mining workflows and results for two exemplars—depression and type 2 diabetes, arguing for translating online shoppers' historical query logs and purchase records to lifestyle-oriented data profiling which is akin to health checkups. Hopefully, unobtrusive chronic disease surveillance via e-commerce sites may soon meet consenting individuals in the digital space they already inhabit.

*Contributions and Implications*

The core contribution of this paper lies in that it links online shopping behaviors, a sound proxy for personal lifestyle choices, with chronic disease prevention, a public health challenge of paramount importance. In the two case studies, the discovered lifestyle risk factors are like fresh blood to research communities, including a great wealth of product categories and buyer personas, most of which exhibit reasonable consistency with the determinants and consequences of depression and type 2 diabetes. Further, these lifestyle risk factors manifest promising predictive power to serve as a scalable front-line alarm offering initial detection of depression and type 2 diabetes to give data-driven decision supports to medical practices. Our experimental results suggest that online shopping behaviors documented in longitudinal query logs and purchase records should be integrated into current modalities for lifestyle-oriented data profiling, especially in today's digital era when social-, mobile-, and local-friendly e-commerce marketing are penetrating people's everyday lives far more easily and profoundly[53]. Along with the continuous improvement of data mining technologies, computational behavioral science in assistance with data profiling for lifestyles may become a dominant methodological paradigm for chronic disease prevention.

This paper also puts forward a critical concern about online shoppers' privacy breaches and informed consent for sharing personal data[54]. Realistically, some people, especially those who have to pay for their health insurance, can be unwilling to share their data for fear of disclosing health issues to any third party. Therefore, health administrators and policymakers should take special care to establish an ethical and supervised information exchange between healthcare networks and e-commerce systems, where online shopping behaviors are supposed to be considered protected health information subject to strict accessibility guidelines. Meanwhile, online shoppers should fully understand the secondary use of their historical query logs and purchase records—what will happen to their data, how their data will be used, and with whom their data will be shared—and maintain autonomous rights in their own healthcare decision-making. As health informatics progresses, unobtrusive chronic disease surveillance via e-commerce sites can be extended further to combine with other digital platforms that mirror personal lifestyle choices, such as social media sites like Twitter and Facebook, to improve health screening and help consenting individuals auto-complete self-report inventories to be assessed by medical experts in case of need.

Last but not least, this paper initiates a new perspective for the social responsibility fulfillment of online retailers in terms of public health—how to run a trade-off between user experience and user wellness. Obviously, e-commerce platforms can employ advanced data mining technologies to customize product recommendations to easily satisfy what online shoppers desire. However, in the long term, some product recommendations driven by customer preferences may irreparably impair online shoppers' physical abilities, especially for the vulnerable with chronic disease potentials. For example, it will cause serious health issues and even premature deaths if online retailers continue advertising sugary cereals and sweetened drinks to "sweet-toothed" online shoppers who have already been informed of a high chance of heart disease. Nevertheless, it is not in the interest of online retailers to stop such harmful advertisement—we have witnessed this in many cases with tobacco, formula milk for babies, etc. Necessary government interventions, such as requiring health warnings to be delivered before e-marketing, should be taken to achieve a win-win situation—online shoppers improve health security, while e-commerce companies increase sustainable profitability.

*Limitation and Future Work*

The findings of this paper should be interpreted with caution due to the following limitation. In this study, whether online shoppers were labeled as "positive" (i.e., diagnosed with a particular chronic disease) or not entirely depended on the presence or absence of related prescription drugs in their purchase records. This practice could be problematic when so-called "positive" online shoppers searched for and bought prescription drugs just for their family members. Similarly, online shoppers who had never made such queries and purchases did not necessarily mean that they were by no means "positive" because they still could receive prescription drugs from offline pharmacies. Note that there are diabetes cases that are controlled through diet and physical activity, and there are off-the-counter (OTC) medicines and non-pharmacologic treatments for depression. Furthermore, despite the increasing cases of type 2 diabetes in China (and elsewhere), the prevalence among teenagers and young adults is still relatively small. There is undeniable misinformation in and about the regression analysis of type 2 diabetes—some diabetes-related medications purchased by "positive" online shoppers at a young age can be used to treat other illnesses. For example, the association with haircare products, wigs, etc. in younger females may be due to the use of metformin as a treatment for polycystic ovary syndrome (PCOS)—hair loss is a symptom of PCOS. One more concern, from the statistical perspective, lies in that some discussion points may go a little beyond what the data tell us, such as (i) supposing that all purchases are made for online shoppers themselves, and (ii) making assumptions about online shoppers' circumstances and link this to a very heteronormative view of marriage and children. However, most of the findings based on such inferences can echo prior studies to a certain degree, and we have made an ambitious attempt to explore the possibility of utilizing online shopping history to explain and understand social phenomena.

This paper does not intend to compete with existing healthcare boards, but fills glaring gaps, from the individual level—empowering digital visitors and residents through near-real-time risk estimates for chronic diseases and situational awareness of lifestyle risk factors affecting their behavioral inertia, to the management level—contextualizing chronic disease prevention relative to evolving landscapes of the ongoing digitalization of daily life. In future work, an appealing direction would be to adopt medical diagnoses, such as the International Classification of Diseases (ICD) codes from consenting individuals[55], as the ground truth for health status assessment for online shoppers, in order to improve the representativeness of "positive" instances of chronic diseases. However, the ethical issues associated with informed consent for data sharing should be reiterated—potential participants need to be informed of what would happen to their data, how their data would be used, and with whom their data would be shared. Another promising line of research would be to incorporate advanced technologies such as deep learning into current data mining workflows to reach fancier knowledge discovery. In addition, we would also like to generalize the proposed approach to investigate as many chronic diseases as possible to expand the horizon of digital health literacy.

## Data Sharing

Since Alibaba takes concerns of data privacy seriously, we would like to emphasize that none of the query logs and purchase records in this research's database permits specific identification with a particular individual, and that the database retains no information about the identity, IP address, or specific physical location of any user. At Alibaba, the query logs and purchase records are considered user's privacy and cannot be shared. However, for the discovered lifestyle risk factors, we can share their means and standard deviations (for ordinal coding) or categorical distributions (for nominal coding) with respect to depressed, diabetic, and control users. The data details are at https://doi.org/10.5281/zenodo.4722474.

**Declaration of Conflicting Interests:** The authors declared no potential conflicts of interest with respect to the research, authorship, and publication of this paper.

**Funding:** YW was supported by the Fundamental Research Funds for the Central Universities under Grant No. DUT21RC(3)068.

**Ethical Approval:** This research was conducted under the permission of Alibaba.com User

Agreements. Moreover, this research was reviewed by the Legal Counsel at Alibaba Group (Process ID: 8721529827) and the Institutional Review Board at Indiana University Bloomington (Protocol #: 10521).

**Contributorship:** YW, XL, and KB designed research; YW and YJ performed data analysis and visualization; YW, JL, CS, and LS performed data collection and cleaning; YW took the lead in writing the paper. All authors reviewed and edited the manuscript and approved the final version of the paper.

**Acknowledgements:** We would like to thank Alibaba Group for providing this research with the query logs and purchase records from 15 million online shoppers within a year span from January 1 to December 31, 2018, and editors and reviewers for their thoughtful comments and suggestions.

# References

1. Raghupathi W and Raghupathi V. An empirical study of chronic diseases in the United States: a visual analytics approach to public health. *Int J Environ Res Public Health* 2018; 15(3): Article 431.
2. Wang L, Kong L, Wu F, et al. Preventing chronic diseases in China. *The Lancet* 2005; 366(9499): 1821-1824.
3. Suhrcke M, Nugent RA, Stuckler D, et al. Cost-effectiveness of interventions to prevent chronic diseases. In: *Chronic Disease: An Economic Perspective*. London, UK: Oxford Health Alliance, 2006, pp.40-47.
4. Chien SY, Chuang MC, Chen I, et al. Primary drivers of willingness to continue to participate in community-based health screening for chronic diseases. *Int J Environ Res Public Health* 2019; 16(9): Article 1645.
5. National Research Council (US) Committee on Diet and Health. Impact of dietary patterns on chronic diseases. In: *Diet and Health: Implications for Reducing Chronic Disease Risk*. Washington, DC: National Academies Press (US), 1989, pp.527-648.

6. Asaria P, Chisholm D, Mathers C, et al. Chronic disease prevention: health effects and financial costs of strategies to reduce salt intake and control tobacco use. *The Lancet* 2007; 370(9604): 2044-2053.

7. Booth FW, Roberts CK and Laye MJ. Lack of exercise is a major cause of chronic diseases. *Compr Physiol* 2011; 2(2): 1143-1211.

8. Shield KD, Parry C and Rehm J. Chronic diseases and conditions related to alcohol use. *Alcohol Res* 2014; 35(2): 155.

9. Snyder T and Byrd G. The internet of everything. *IEEE Computer Architecture Letters* 2017; 50(06): 8-9.

10. Ravizza S, Huschto T, Adamov A, et al. Predicting the early risk of chronic kidney disease in patients with diabetes using real-world data. *Nat Med* 2019; 25(1): 57-59.

11. De Choudhury M, Gamon M, Counts S, et al. Predicting depression via social media. In: *Proceedings of the 7th International AAAI Conference on Web and Social Media*, Cambridge, MA, 8-11 July 2013, pp.128-137.

12. Shen JH and Rudzicz F. Detecting anxiety through reddit. In: *Proceedings of the 4th Workshop on Computational Linguistics and Clinical Psychology—From Linguistic Signal to Clinical Reality*, Vancouver, Canada, 3 August 2017, pp.58-65.

13. Reece AG and Danforth CM. Instagram photos reveal predictive markers of depression. *EPJ Data Science* 2017; 6: Article 15.

14. Eichstaedt JC, Smith RJ, Merchant RM, et al. Facebook language predicts depression in medical records. *Proc Natl Acad Sci U S A* 2018; 115(44): 11203-11208.

15. Kibin. The impact of online shopping on the lifestyle of people in our modern society. http://www.kibin.com/essay-examples/the-impact-of-online-shopping-on-the-lifestyle-of-people-in-our-modern-society-zeYeyiIq (2021, accessed 23 November 2021)

16. Coppola D. Global number of digital buyers 2014-2021. https://www.statista.com/statistics/251666/number-of-digital-buyers-worldwide/ (2021, accessed 5 April 2021)

17. Chevalier S. Retail e-commerce sales worldwide from 2014 to 2024. https://www.statista.com/statistics/379046/worldwide-retail-e-commerce-sales/ (2021, accessed 24 November 2021)

18. Cain MM, Sarasohn-Kahn J and Wayne JC. Who are health e-people? A segmentation of online health consumers. In: *Health e-People: The Online Consumer Experience*. Auckland, CA: California HealthCare Foundation, 2000, pp.9-12.
19. Clark D and Weir C. 2019: China to surpass US in total retail sales. https://www.emarketer.com/newsroom/index.php/2019-china-to-surpass-us-in-total-retail-sales/ (2019, accessed 15 January 2021)
20. Blystone D. Understanding the Alibaba business model. https://www.investopedia.com/articles/investing/062315/understanding-alibabas-business-model.asp (2021, accessed 20 January 2021)
21. Sarris J, Thomson R, Hargraves F, et al. Multiple lifestyle factors and depressed mood: a cross-sectional and longitudinal analysis of the UK Biobank (N=84,860). *BMC Med* 2020; 18: 354.
22. Reddy PH. Can diabetes be controlled by lifestyle activities?. *Curr Res Diabetes Obes J* 2017; 1(4): Article 555568.
23. Qin X, Wang S and Hsieh CR. The prevalence of depression and depressive symptoms among adults in China: estimation based on a National Household Survey. *China Economic Review* 2018; 51: 271-282.
24. Que J, Lu L and Shi L. Development and challenges of mental health in China. *Gen Psychiatr* 2019; 32(1): Article e100053.
25. Ma RC. Epidemiology of diabetes and diabetic complications in China. *Diabetologia* 2018; 61(6): 1249-1260.
26. Wang L, Gao P, Zhang M, et al. Prevalence and ethnic pattern of diabetes and prediabetes in China in 2013. *JAMA* 2017; 317(24): 2515-2523.
27. World Health Organization. Global and regional estimates of prevalence: depressive disorders. In: *Depression and Other Common Mental Disorders: Global Health Estimates*. Geneva, Switzerland: WHO Document Production Services, 2017, pp.8-9.
28. Wu L. Rate of diabetes in China "explosive". https://www.who.int/china/news/detail/06-04-2016-rate-of-diabetes-in-china-explosive- (2016, accessed 15 January 2021)
29. McDonald JH. Multiple logistic regression. In: *Handbook of Biological Statistics (3rd Ed.)*. Baltimore, MD: Sparky House Publishing, 2014, pp.247-253.

30. Chen H, Cohen P and Chen S. How big is a big odds ratio? Interpreting the magnitudes of odds ratios in epidemiological studies. *Commun Stat Simul Comput* 2010; 39(4): 860-864.

31. Ma Y. Distribution of online buyers in China in 2019, by age group. https://www.statista.com/statistics/1172011/china-age-group-distribution-of-online-shoppers/ (2021, accessed 25 November 2021)

32. Coppola D. Distribution of digital buyers in the United States as of February 2020, by age group. https://www.statista.com/statistics/469184/us-digital-buyer-share-age-group/ (2021, accessed 25 November 2021)

33. Benjamini Y and Hochberg Y. Controlling the false discovery rate: a practical and powerful approach to multiple testing. *J R Stat Soc Series B Stat Methodol* 1995; 57(1): 289-300.

34. Cortes C and Vapnik V. Support-vector networks. *Machine Learning*, 1995; 20(3): 273-297.

35. Swets JA. Measuring the accuracy of diagnostic systems. *Science* 1988; 240(4857): 1285-1293.

36. Zisook S, Lesser I, Stewart JW, et al. Effect of age at onset on the course of major depressive disorder. *Am J Psychiatry* 2007; 164(10): 1539-1546.

37. Elmer T and Stadtfeld C. Depressive symptoms are associated with social isolation in face-to-face interaction networks. *Sci Rep* 2020; 10(1): Article 1444.

38. Kim HK and McKenry PC. The relationship between marriage and psychological well-being: A longitudinal analysis. *Journal of Family Issues* 2020; 23(8): 885-911.

39. Pedrelli P, Shapero B, Archibald A, et al. Alcohol use and depression during adolescence and young adulthood: a summary and interpretation of mixed findings. *Curr Addict Rep* 2016; 3(1): 91-97.

40. Gokalp H. Psychosocial aspects of hair loss. In: *Hair and Scalp Disorders*. London, UK: IntechOpen, 2017, pp.239-252.

41. Simon GE, Ludman EJ, Linde JA, et al. Association between obesity and depression in middle-aged women. *Gen Hosp Psychiatry* 2008; 30(1): 32-39.

42. West A. Mental health and money—The Anxiety and Depression Association of America (ADAA) weighs in on how financial stress affects your well-being. https://www.badcredit.org/news/adaa-weighs-in-on-how-financial-stress-affects-your-well-being/ (2018, accessed 8 March 2021)

43. Huebschmann AG, Huxley RR, Kohrt WM, et al. Sex differences in the burden of type 2 diabetes and cardiovascular risk across the life course. *Diabetologia* 2019; 62(10): 1761-1772.

44. Sami W, Ansari T, Butt NS, et al. Effect of diet on type 2 diabetes mellitus: A review. *International Journal of Health Sciences* 2017; 11(2): Article 65.

45. Hsu CC, Lee CH, Wahlqvist ML, et al. Poverty increases type 2 diabetes incidence and inequality of care despite universal health coverage. *Diabetes Care* 2012; 35(11): 2286-2292.

46. Attard SM, Herring AH, Mayer-Davis EJ, et al. Multilevel examination of diabetes in modernising China: what elements of urbanisation are most associated with diabetes?. *Diabetologia*, 2012; 55(12): 3182-3192.

47. Lazar MA. How obesity causes diabetes: not a tall tale. *Science* 2005; 307(5708): 373-375.

48. Cornelis MC, Chiuve SE, Glymour MM, et al. Bachelors, divorcees, and widowers: does marriage protect men from type 2 diabetes?. *PLoS One* 2014; 9(9): Article e106720.

49. Trinh NHT, Youn SJ, Sousa J, et al. Using electronic medical records to determine the diagnosis of clinical depression. *Int J Med Inform* 2011; 80(7): 533-540.

50. Gao WG, Dong YH, Pang ZC, et al. A simple Chinese risk score for undiagnosed diabetes. *Diabet Med* 2010; 27(3): 274-281.

51. Hariri S, Yoon PW, Qureshi N, et al. Family history of type 2 diabetes: a population-based screening tool for prevention?. *Genet Med* 2006; 8(2): 102-108.

52. Cheung BM and Li C. Diabetes and hypertension: is there a common metabolic pathway?. *Curr Atheroscler Rep* 2012; 14(2): 160-166.

53. Laudon KC and Traver CG. Social, mobile, and local marketing. In: *E-commerce: Business, Technology, Society (12th Ed.)*. Chicago, IL: RR Donnelley, 2016, pp.460-529.

54. Eaton I and McNett M. Protecting the data: security and privacy. In: *Data for Nurses: Understanding and Using Data to Optimize Care Delivery in Hospitals and Health Systems (1st Ed.)*. Cambridge, MA: Academic Press, 2019, pp.87-99.

55. O'malley KJ, Cook KF, Price MD, et al. Measuring diagnoses: ICD code accuracy. *Health Services Research* 2005; 40(5p2): 1620-1639.

Supplementary Information for

# Leveraging online shopping behaviors as a proxy for personal lifestyle choices: New insights into chronic disease prevention literacy

**Content**

# Appendix A. Health Status Assessment

**Table A.1.** Prescription drugs used to assess online shoppers' health status, found in Alibaba users' purchase records and confirmed by several domain experts together. Here, the generic name of each prescription drug is listed for simplicity. Note that the Anatomical Therapeutic Chemical (ATC) is an internationally accepted classification system for medicines maintained by the World Health Organization (WHO). The WHO assigns ATC codes to all active substances contained in medicines based on the therapeutic indication for the medicine.

| Chronic Disease | Prescription Drug | ATC Code | Chronic Disease | Prescription Drug | ATC Code |
| --- | --- | --- | --- | --- | --- |
| Depression | Agomelatine | N06AX22 | Type 2 Diabetes | Acarbose | A10BF01 |
|  | Amitriptyline | N06AA09 |  | Canagliflozin | A10BK02 |
|  | Bupropion | N06AX12 |  | Dapagliflozin | A10BK01 |
|  | Citalopram | N06AB04 |  | Empagliflozin | A10BK03 |
|  | Clomipramine | N06AA04 |  | Exenatide | A10BJ01 |
|  | Doxepin | D04AX01 |  | Glibenclamide | A10BB01 |
|  | Duloxetine | N06AX21 |  | Gliclazide | A10BB09 |
|  | Escitalopram | N06AB10 |  | Glimepiride | A10BB12 |
|  | Fluoxetine | N06AB03 |  | Glipizide | A10BB07 |
|  | Fluvoxamine | N06AB08 |  | Gliquidone | A10BB08 |
|  | Melitracen | N06AA14 |  | Insulin | A10AD01 |
|  | Mirtazapine | N06AX11 |  | Linagliptin | A10BH05 |
|  | Paroxetine | N06AB05 |  | Liraglutide | A10BJ02 |
|  | Sertraline | N06AB06 |  | Metformin | A10BA02 |
|  | Trazodone | N06AX05 |  | Miglitol | A10BF02 |
|  | Venlafaxine | N06AX16 |  | Nateglinide | A10BX03 |
|  |  |  |  | Pioglitazone | A10BG03 |
|  |  |  |  | Repaglinide | A10BX02 |
|  |  |  |  | Rosiglitazone | A10BG02 |
|  |  |  |  | Saxagliptin | A10BH03 |
|  |  |  |  | Sitagliptin | A10BH01 |
|  |  |  |  | Voglibose | A10BF03 |

# Appendix B. Difference in Monthly Query Logs and Purchase Records Among Online Shoppers of Different Sex and Age

**Table B.1.** Data statistics. Here, 10000 Alibaba users were randomly sampled for analysis purposes.

| Sex | Age | | | | | | Total |
|---|---|---|---|---|---|---|---|
| | **15-24** | **25-34** | **35-44** | **45-54** | **55-64** | **≥65** | |
| Female | 1414 | 2433 | 1339 | 450 | 27 | 5 | 5668 |
| Male | 1141 | 1719 | 1059 | 378 | 35 | 0 | 4332 |
| Total | 2555 | 4152 | 2398 | 828 | 62 | 5 | 10000 |

**Table B.2.** Numbers of query logs per month (mean ± standard deviation) for online shoppers of different sex and age.

| Sex | Age | | | | |
|---|---|---|---|---|---|
| | **15-24** | **25-34** | **35-44** | **45-54** | **55-64** |
| Female | 69.18±65.12 | 48.16±49.00 | 42.35±51.43 | 36.49±42.66 | 22.04±19.56 |
| Male | 40.31±42.20 | 38.68±41.80 | 39.13±45.29 | 38.95±40.91 | 34.51±46.23 |

**Table B.3.** Numbers of purchase records per month (mean ± standard deviation) for online shoppers of different sex and age.

| Sex | Age | | | | |
|---|---|---|---|---|---|
| | **15-24** | **25-34** | **35-44** | **45-54** | **55-64** |
| Female | 15.49±10.77 | 16.27±13.85 | 13.68±13.76 | 10.91±10.70 | 10.38±11.24 |
| Male | 9.35±8.06 | 11.21±11.46 | 11.08±10.76 | 10.83±10.99 | 14.27±11.34 |

**Table B.4.** Scheirer-Ray-Hare analysis of variance for numbers of query logs per month for online shoppers of different sex and age.

| SV | SS | df | MS | H | p | Sig. |
|---|---|---|---|---|---|---|
| Sex | 45722727.10 | 1 | | 5.60 | 0.02 | Yes |
| Age | 1427434402.99 | 4 | | 174.70 | 0 | Yes |
| Sex × Age | 808029727.08 | 4 | | 98.89 | 0 | Yes |
| Error | 79376103144.26 | 9985 | | | | |
| Total | 81657290001.44 | 9994 | 8170631.38 | | | |

*SV*: source of variance

*SS*: sum of squares

*MS*: mean squares

**Table B.5.** Scheirer-Ray-Hare analysis of variance for numbers of purchase records per month for online shoppers of different sex and age.

| SV | SS | df | MS | H | p | Sig. |
|---|---|---|---|---|---|---|
| Sex | 90699574.12 | 1 | | 11.42 | 7.25E-04 | Yes |
| Age | 577055091.73 | 4 | | 72.68 | 1.00E-14 | Yes |
| Sex × Age | 1038885216.97 | 4 | | 130.85 | 0 | Yes |
| Error | 77643031162.96 | 9985 | | | | |
| Total | 79349671045.78 | 9994 | 7939730.94 | | | |

*SV*: source of variance

*SS*: sum of squares

*MS*: mean squares

# Appendix C. Feature Selection

**Table C.1.** Lifestyle features selected for determining associations between online shoppers' past lifestyle choices and whether they suffer from depression ($N$=59140) when demographics are controlled for. The Pearson's Chi-squared test is used to examine whether a lifestyle feature is independent of the prevalence and distribution of depression among online shoppers. A humble significance threshold $p<0.1$ is used for this independence test to avoid missing potentially important lifestyle features. Here, each listed lifestyle feature is employed by at least one subgroup analysis.

*$p<0.1$, **$p<0.05$, ***$p<0.01$, ****$p<0.001$

| | Subsample (Sex, Age, and Size) | | | | | | | |
|---|---|---|---|---|---|---|---|---|
| | Female | | | | Male | | | |
| Lifestyle Feature | 15-24 | 25-34 | 35-44 | 45-54 | 15-24 | 25-34 | 35-44 | 45-54 |
| | $n$=5660 | $n$=6500 | $n$=6200 | $n$=3900 | $n$=7400 | $n$=13840 | $n$=10300 | $n$=5340 |
| 3C Digital Accessories | | | | * | | | | |
| Added & Redeemed Coupons | | | | | | ** | | |
| Air Conditioner Preference | | | | | | | ** | |
| Alcohol Preference | **** | | **** | **** | **** | | ** | |
| Audio & Visual Equipment | *** | | | | **** | * | | |
| Auto Parts & Accessories | | | | ** | | ** | ** | |
| Bags & Suitcases | | | | ** | | | | |
| Beauty & Cosmetics Preference | | | ** | | | ** | *** | |
| Bedding Comforters & Sets | | | *** | ** | | | | |
| Beverage Preference | *** | **** | ** | **** | **** | | | |
| Body Wash Preference | ** | | | | | | * | |

| Lifestyle Feature | Subsample (Sex, Age, and Size) | | | | | | | |
|---|---|---|---|---|---|---|---|---|
| | Female | | | | Male | | | |
| | 15-24 | 25-34 | 35-44 | 45-54 | 15-24 | 25-34 | 35-44 | 45-54 |
| | *n*=5660 | *n*=6500 | *n*=6200 | *n*=3900 | *n*=7400 | *n*=13840 | *n*=10300 | *n*=5340 |
| Body Weight | **** | ** | **** | **** | ** | | ** | ** |
| Bodycare Products | ** | | *** | * | ** | ** | | |
| Books, Magazines & Newspapers | **** | *** | | | **** | **** | *** | |
| Camera Phone Preference | * | * | ** | | | | | * |
| Candies | | | | *** | * | | | |
| Candy Preference | * | | *** | | | | | |
| Cereals, Dried Foods & Condiments | | *** | **** | | ** | ** | **** | **** |
| Children's Clothing | | *** | * | | | **** | ** | |
| Children's Shoes | | ** | | | | | | |
| Children's Toys | | * | | **** | | | | |
| Clothing Accessories | **** | | | | ** | | | *** |
| Coffee, Oatmeal & Powdered Drink Mixes | **** | **** | **** | **** | **** | | | |
| Coffee Preference | *** | **** | ** | | ** | | | |
| Computer Hardware & Peripherals | | | | | * | | | |
| Cooking Utensils | | * | * | | | | | |
| Cosmetics & Fragrances & Beauty Tools | * | | | | ** | | | |
| Credit Score | ** | **** | **** | **** | * | **** | ** | **** |
| Diaper Preference | * | | | | | | | |
| Down Jacket Preference | | | **** | | | ** | | |

| Lifestyle Feature | Subsample (Sex, Age, and Size) | | | | | | | |
|---|---|---|---|---|---|---|---|---|
| | Female | | | | Male | | | |
| | 15-24 | 25-34 | 35-44 | 45-54 | 15-24 | 25-34 | 35-44 | 45-54 |
| | *n*=5660 | *n*=6500 | *n*=6200 | *n*=3900 | *n*=7400 | *n*=13840 | *n*=10300 | *n*=5340 |
| E-Books & Stationery | **** | ** | | *** | *** | * | | |
| Energy-Hungry Appliance Preference | | | | * | | | | * |
| Fashion Ornaments | **** | ** | | | | | | |
| Financial Status | **** | **** | **** | **** | **** | **** | **** | **** |
| Fitness & Massage Tools | | *** | **** | **** | ** | **** | | |
| Flower Express | | | | | | ** | | |
| Foundation Materials | | | | | | *** | | ** |
| Glasses & Smoking Accessories | ** | | | | | | | |
| Haircare Products & Wigs | **** | **** | ** | *** | | | | |
| Haircare Services | **** | | | ** | | | | |
| High-Capacity-Battery Phone Preference | | | | | | * | | |
| Home Decoration Preference | *** | | **** | ** | | ** | **** | ** |
| Home Furniture | | | | *** | | | | |
| Home Healthcare Supplies | **** | **** | **** | **** | **** | **** | **** | **** |
| Home Improvements | ** | | | **** | | | | |
| Home Storage & Organization Products | | | | ** | **** | | | |
| Home Textiles | *** | | | ** | * | | | |
| Household & Personal Hygiene Tools | | | **** | | *** | | | |
| Income Level | | ** | **** | ** | | **** | **** | ** |

| Lifestyle Feature | Subsample (Sex, Age, and Size) | | | | | | | |
|---|---|---|---|---|---|---|---|---|
| | Female | | | | Male | | | |
| | 15-24 | 25-34 | 35-44 | 45-54 | 15-24 | 25-34 | 35-44 | 45-54 |
| | *n*=5660 | *n*=6500 | *n*=6200 | *n*=3900 | *n*=7400 | *n*=13840 | *n*=10300 | *n*=5340 |
| Kitchen Appliances | | | * | * | | | | |
| Large-Screen Phone Preference | | | | | * | | ** | |
| Large-Storage-Space Phone Preference | | | ** | | ** | | | |
| Meats, Seafoods & Vegetables | *** | *** | *** | *** | | **** | **** | *** |
| Megapixel Camera Preference | | ** | ** | | | | | |
| Membership Level | *** | **** | **** | **** | **** | **** | *** | ** |
| Men's Clothing | **** | ** | | | *** | | | * |
| Men's Clothing Preference | | ** | * | | *** | | | * |
| Mid-Range Phone Preference | **** | *** | *** | *** | **** | **** | **** | |
| Mini Household Appliances | | * | | | | | | |
| Miscellaneous Household Items | ** | | *** | | *** | * | | |
| Outdoor Gear | | * | * | *** | * | | | |
| Pajamas & Underwear | | | *** | | | | * | |
| Pet Food Preference | *** | | | | | | | |
| Phone Carrier Preference | | | | | | | | * |
| Phone Color Preference | ** | | | * | * | | * | |
| Phone Expenses | *** | | | | | | | |
| Phone Preference | * | *** | ** | | | | | |
| Phone Purchase Price | | **** | | | *** | **** | | ** |

| Lifestyle Feature | Subsample (Sex, Age, and Size) | | | | | | | |
|---|---|---|---|---|---|---|---|---|
| | Female | | | | Male | | | |
| | 15-24 | 25-34 | 35-44 | 45-54 | 15-24 | 25-34 | 35-44 | 45-54 |
| | *n*=5660 | *n*=6500 | *n*=6200 | *n*=3900 | *n*=7400 | *n*=13840 | *n*=10300 | *n*=5340 |
| Postage | **** | **** | | | | | | |
| Posted Positive Reviews | **** | ** | | | ** | | | |
| Purchase Amount | **** | **** | **** | **** | **** | **** | **** | * |
| Purchase Frequency | **** | *** | | **** | **** | **** | **** | **** |
| Purchased Product Categories | | | | | | | | *** |
| Second-Hand Goods | | | | | **** | | | |
| Skincare Preference | *** | *** | | | | | | |
| Skincare Products | **** | * | **** | | *** | * | | ** |
| Spending Per Purchase | | | ** | | | * | | |
| Sports Preference | ** | | ** | | ** | * | | |
| Sports, Yoga & Gym Articles | | * | | ** | | | | |
| Snacks | *** | *** | **** | *** | **** | ** | **** | * |
| Sneakers | **** | | | | * | | | |
| Sweatpants Preference | | ** | * | | | | | |
| Tableware | *** | | * | | * | | | |
| Tea Beverages | ** | | ** | * | | | | * |
| Tea Preference | | ** | *** | | | ** | | |
| Tier Of City Of Residence | *** | | | ** | | | | |
| Tissue Paper Preference | | | | | ** | | | |

| Lifestyle Feature | Subsample (Sex, Age, and Size) | | | | | | | |
|---|---|---|---|---|---|---|---|---|
| | Female | | | | Male | | | |
| | 15-24 | 25-34 | 35-44 | 45-54 | 15-24 | 25-34 | 35-44 | 45-54 |
| | *n*=5660 | *n*=6500 | *n*=6200 | *n*=3900 | *n*=7400 | *n*=13840 | *n*=10300 | *n*=5340 |
| Toiletries | **** | | **** | | *** | ** | **** | |
| Video Games | | | | | ** | | | |
| Women's Clothing | | ** | * | | | * | ** | |
| Women's Clothing Preference | ** | * | *** | * | | **** | ** | *** |
| Women's Shoes | | | | | | | * | |
| Workout Clothes Preference | | | | | ** | | | |
| Total Selected | 48 | 41 | 45 | 38 | 44 | 34 | 27 | 24 |

**Table C.2.** Lifestyle features selected for determining associations between online shoppers' past lifestyle choices and whether they suffer from type 2 diabetes (*N*=38420) when demographics are controlled for. The Pearson's Chi-squared test is used to examine whether a lifestyle feature is independent of the prevalence and distribution of type 2 diabetes among online shoppers. A humble significance threshold *p*<0.1 is used for this independence test to avoid missing important lifestyle features. Here, each listed lifestyle feature is employed by at least one subgroup analysis.
\**p*<0.1, \*\**p*<0.05, \*\*\**p*<0.01, \*\*\*\**p*<0.001

| Lifestyle Feature | Subsample (Sex, Age, and Size) | | | | | | | | | |
|---|---|---|---|---|---|---|---|---|---|---|
| | Female | | | | | Male | | | | |
| | 15-24 *n*=2960 | 25-34 *n*=6400 | 35-44 *n*=4940 | 45-54 *n*=2870 | 55-64 *n*=1160 | 15-24 *n*=1600 | 25-34 *n*=5520 | 35-44 *n*=5930 | 45-54 *n*=5040 | 55-64 *n*=2000 |
| 3C Digital Accessories | | **** | **** | * | | *** | **** | **** | ** | |
| Added & Redeemed Coupons | | **** | **** | ** | | | *** | | | |
| Air Conditioner Preference | | | | | | | ** | | | |
| Alcohol Preference | ** | **** | | | | * | * | ** | | |
| Audio & Visual Equipment | ** | * | * | | | | *** | **** | | |
| Baby Formula Preference | | **** | | | | | | | | |
| Babycare Products | | ** | **** | | | | | | | |
| Bags & Suitcases | **** | | | | | | | | | |
| Beauty & Cosmetics Preference | | | | | | | | | *** | |
| Bedding Comforters & Sets | | **** | *** | | | | *** | | | |
| Beverage Preference | *** | **** | **** | | | **** | **** | * | | |
| Body Wash Preference | **** | **** | | | | | | | | |
| Body Weight | ** | **** | **** | *** | | | * | **** | **** | |

| Lifestyle Feature | Subsample (Sex, Age, and Size) | | | | | | | | | |
|---|---|---|---|---|---|---|---|---|---|---|
| | Female | | | | | Male | | | | |
| | 15-24 | 25-34 | 35-44 | 45-54 | 55-64 | 15-24 | 25-34 | 35-44 | 45-54 | 55-64 |
| | $n$=2960 | $n$=6400 | $n$=4940 | $n$=2870 | $n$=1160 | $n$=1600 | $n$=5520 | $n$=5930 | $n$=5040 | $n$=2000 |
| Bodycare Products | *** | **** | **** | **** | | ** | **** | *** | | |
| Books, Magazines & Newspapers | * | | | | | **** | | | | |
| Candies | | **** | **** | ** | | ** | | | | |
| Cereals, Dried Foods & Condiments | **** | **** | **** | **** | ** | **** | **** | **** | **** | **** |
| Children's Clothing | | | ** | | | | | | * | |
| Children's Toys | | | *** | | | | | | | |
| Chinese Tonics | | | **** | **** | *** | | | | **** | *** |
| Clothing Accessories | | *** | *** | | | | * | | | |
| Coffee, Oatmeal & Powdered Drink Mixes | **** | **** | **** | *** | | *** | **** | | | **** |
| Coffee Preference | **** | **** | * | | | | | * | | |
| Cooking Utensils | ** | **** | **** | **** | *** | | *** | | | |
| Cosmetics & Fragrances & Beauty Tools | | *** | *** | | | * | | | | |
| Credit Score | | **** | **** | **** | *** | ** | **** | **** | | * |
| Diapers | | * | | | | | | | | |
| Diaper Preference | | ** | *** | | | | | | | |
| Down Jacket Preference | | **** | | | | | | | | |
| E-Books & Stationery | ** | *** | | | | | * | | * | |
| Energy-Hungry Appliance Preference | ** | **** | **** | | | ** | | | | |
| Fashion Ornaments | * | **** | **** | | | | | | | |

| Lifestyle Feature | Subsample (Sex, Age, and Size) | | | | | | | | | |
| --- | --- | --- | --- | --- | --- | --- | --- | --- | --- | --- |
| | Female | | | | | Male | | | | |
| | 15-24 | 25-34 | 35-44 | 45-54 | 55-64 | 15-24 | 25-34 | 35-44 | 45-54 | 55-64 |
| | *n*=2960 | *n*=6400 | *n*=4940 | *n*=2870 | *n*=1160 | *n*=1600 | *n*=5520 | *n*=5930 | *n*=5040 | *n*=2000 |
| Festive Products | | ** | **** | | | | | | | |
| Financial Status | **** | **** | **** | **** | ** | **** | **** | **** | **** | *** |
| Fitness & Massage Tools | | **** | *** | ** | | | * | *** | **** | ** |
| Flower Express | | * | **** | ** | | | *** | | | |
| Foundation Materials | | | | | | | **** | | | |
| Haircare Products & Wigs | ** | **** | **** | | | | | | | |
| Haircare Services | *** | | | ** | | | | | | |
| High-Capacity-Battery Phone Preference | * | **** | * | ** | | | ** | | | |
| Home Decoration Preference | | **** | ** | | | ** | *** | **** | **** | *** |
| Home Decorative Lighting | | | | | | | ** | | | |
| Home Decorative Materials | | | | | | | *** | | * | |
| Home Furniture | | | *** | * | | | | | * | |
| Home Healthcare Supplies | **** | **** | **** | **** | **** | **** | **** | **** | **** | **** |
| Home Improvements | | *** | | | | | * | | *** | |
| Home Storage & Organization Products | ** | *** | *** | | | | | | | |
| Home Textiles | | *** | **** | | | | | | | |
| Household & Personal Hygiene Tools | ** | **** | ** | *** | | | **** | | ** | * |
| Income Level | **** | **** | **** | * | ** | *** | * | * | **** | * |
| Kitchen Appliances | | **** | *** | **** | | ** | **** | ** | | |

| Lifestyle Feature | Subsample (Sex, Age, and Size) | | | | | | | | | |
|---|---|---|---|---|---|---|---|---|---|---|
| | Female | | | | | Male | | | | |
| | 15-24 $n$=2960 | 25-34 $n$=6400 | 35-44 $n$=4940 | 45-54 $n$=2870 | 55-64 $n$=1160 | 15-24 $n$=1600 | 25-34 $n$=5520 | 35-44 $n$=5930 | 45-54 $n$=5040 | 55-64 $n$=2000 |
| Large-Screen Phone Preference | | | | | | | ** | ** | *** | |
| Large-Screen TV Preference | | **** | | | | | | | | |
| Large-Storage-Space Phone Preference | ** | *** | | | | | ** | | | |
| Major Household Appliances | | | | | | | | | | ** |
| Meats, Seafoods & Vegetables | **** | **** | **** | **** | | **** | **** | **** | **** | ** |
| Megapixel Camera Preference | ** | | | | *** | | | | | |
| Membership Level | * | **** | **** | **** | | ** | **** | **** | **** | |
| Men's Clothing | | ** | * | | | | | | | * |
| Men's Clothing Preference | | *** | | | | ** | | | | * |
| Men's Shoes | | | | | | | | | ** | * |
| Mid-Range Phone Preference | ** | **** | **** | ** | | | **** | | **** | |
| Mini Household Appliances | | ** | | | | | | | | |
| Miscellaneous Household Items | *** | **** | **** | | | * | *** | ** | *** | |
| Outdoor Gear | | ** | **** | | | | | ** | *** | |
| Pajamas & Underwear | | **** | * | | | | | | | |
| Pet Foods & Supplies | | | | * | | | | | | |
| Phone Carrier Preference | **** | | | ** | | | *** | ** | | * |
| Phone Expenses | * | ** | | | | | | | | |
| Phone Preference | ** | **** | | ** | ** | | **** | *** | | * |

| Lifestyle Feature | Subsample (Sex, Age, and Size) | | | | | | | | | |
|---|---|---|---|---|---|---|---|---|---|---|
| | Female | | | | | Male | | | | |
| | 15-24 | 25-34 | 35-44 | 45-54 | 55-64 | 15-24 | 25-34 | 35-44 | 45-54 | 55-64 |
| | *n*=2960 | *n*=6400 | *n*=4940 | *n*=2870 | *n*=1160 | *n*=1600 | *n*=5520 | *n*=5930 | *n*=5040 | *n*=2000 |
| Phone Purchase Price | ** | * | ** | | | | | | | |
| Placed Orders | **** | | | | | **** | **** | | ** | |
| Postage | **** | **** | | | | | | | | |
| Posted Positive Reviews | **** | | | | | *** | | | *** | |
| Purchase Amount | **** | **** | **** | **** | | ** | **** | **** | | |
| Purchase Frequency | | | | | * | | | **** | **** | *** |
| Purchased Product Categories | | | | | | | | | **** | |
| Second-Hand Goods | **** | | | | | **** | | | | |
| Skincare Preference | | | | | | ** | | | | |
| Skincare Products | **** | **** | **** | | | | ** | | | |
| Snacks | *** | **** | **** | *** | | *** | **** | **** | | *** |
| Spending Per Purchase | | **** | **** | | | | **** | **** | | |
| Sports Preference | | ** | ** | *** | | | * | | * | |
| Sports, Yoga & Gym Articles | * | **** | | | | | * | | | |
| Sweatpants Preference | | **** | | | * | | | | | |
| Tableware | **** | **** | * | ** | | | | | | |
| Tea | | | | | | | | **** | *** | |
| Tea Beverages | ** | **** | **** | **** | | ** | **** | **** | **** | |
| Tier Of City Of Residence | ** | **** | **** | **** | **** | * | **** | **** | **** | **** |

| | Subsample (Sex, Age, and Size) | | | | | | | | | |
|---|---|---|---|---|---|---|---|---|---|---|
| **Lifestyle Feature** | **Female** | | | | | **Male** | | | | |
| | 15-24 | 25-34 | 35-44 | 45-54 | 55-64 | 15-24 | 25-34 | 35-44 | 45-54 | 55-64 |
| | $n$=2960 | $n$=6400 | $n$=4940 | $n$=2870 | $n$=1160 | $n$=1600 | $n$=5520 | $n$=5930 | $n$=5040 | $n$=2000 |
| Tissue Paper Preference | | **** | *** | | | **** | **** | *** | | |
| Toiletries | *** | **** | **** | **** | * | **** | **** | **** | *** | |
| Video Games | | | | | | ** | | | | |
| Women's Clothing | | *** | | | | ** | | **** | * | |
| Women's Clothing Preference | | | ** | | *** | *** | | **** | **** | **** |
| Women's Shoes | | | | ** | | | | | | |
| Workout Clothes Preference | | **** | | | | | | | | |
| Total Selected | 45 | 68 | 55 | 34 | 14 | 33 | 48 | 36 | 31 | 20 |

# Appendix D. Parameter Optimization

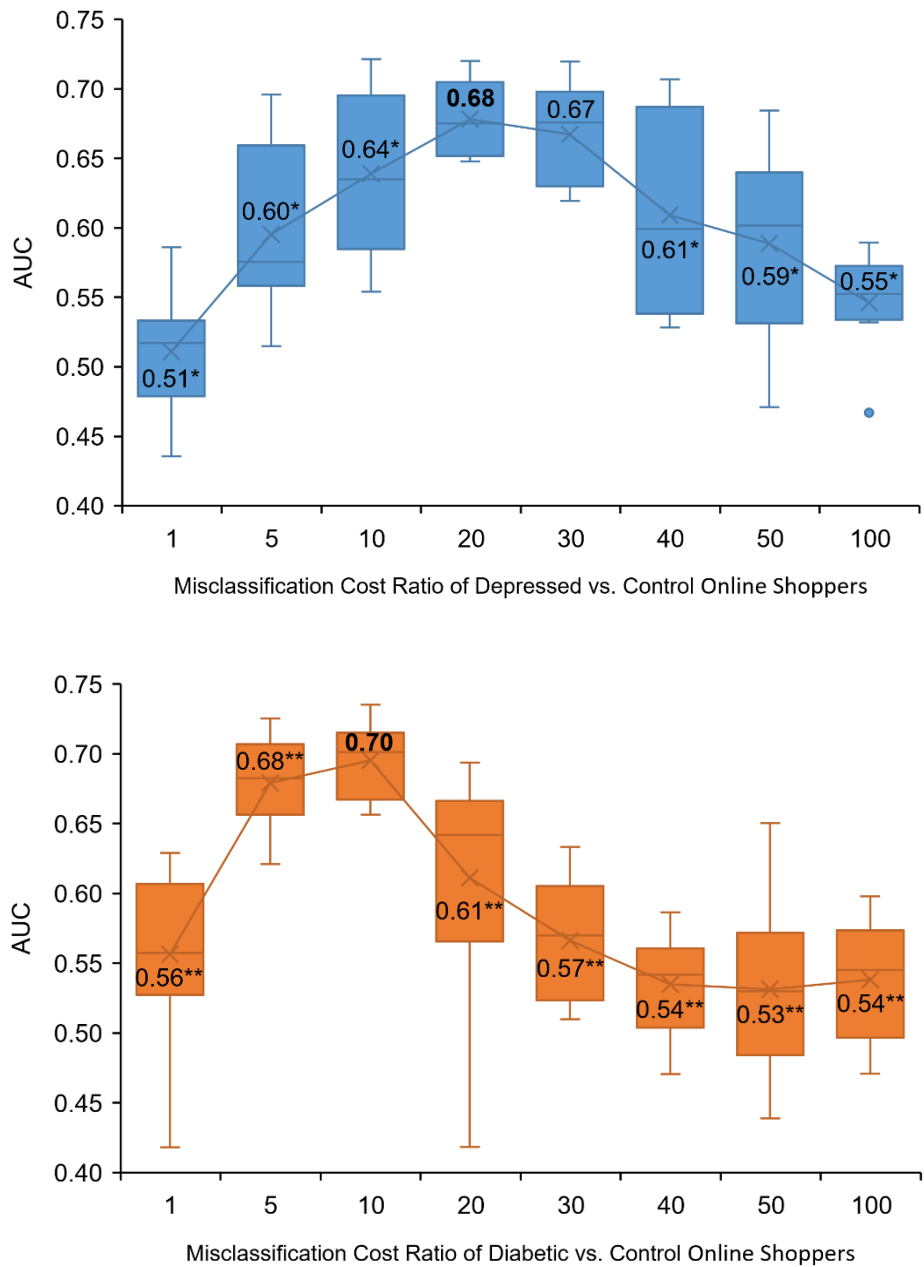

**Figure D.1.** Misclassification cost optimization for online shopping behaviors-based predictive classifiers (via SVM) for depression (*Top*, *N*=8) and type 2 diabetes (*Bottom*, *N*=10) when demographics are controlled for. Statistical analysis is conducted by the Wilcoxon signed-rank test, in comparison with the optimal misclassification cost (marked in bold).
*$p<0.05$, **$p<0.01$
AUC: area under the receiver operating characteristic curve

# Appendix E. Regression Output

**Table E.1.** Multiple logistic regression (using the enter method) for determining associations between online shoppers' past lifestyle choices and whether they suffer from depression ($N$=59140) when controlling for demographics.

| Subsample | | | Omnibus Tests of Model Coefficients | | | Model Summary | | | Hosmer and Lemeshow Test | | |
|---|---|---|---|---|---|---|---|---|---|---|---|
| Sex | Age | Size | $\chi^2$ | $df$ | $p$ | -2 Log Likelihood | Cox & Snell $R^2$ | Nagelkerke $R^2$ | $\chi^2$ | $df$ | $p$ |
| Female | 15-24 | 5,660 | 435.037 | 151 | 0.000 | 1812.156 | 0.074 | 0.226 | 8.616 | 8 | 0.376 |
| | 25-34 | 6,500 | 313.642 | 101 | 0.000 | 2267.056 | 0.047 | 0.144 | 6.239 | 8 | 0.620 |
| | 35-44 | 6,200 | 299.706 | 161 | 0.000 | 2161.883 | 0.047 | 0.144 | 3.475 | 8 | 0.901 |
| | 45-54 | 3,900 | 207.887 | 95 | 0.000 | 1340.531 | 0.052 | 0.158 | 9.068 | 8 | 0.337 |
| Male | 15-24 | 7,400 | 359.866 | 110 | 0.000 | 2578.160 | 0.047 | 0.145 | 8.949 | 8 | 0.347 |
| | 25-34 | 13,840 | 462.581 | 101 | 0.000 | 5032.321 | 0.033 | 0.100 | 13.519 | 8 | 0.095 |
| | 35-44 | 10,300 | 360.253 | 118 | 0.000 | 3729.161 | 0.034 | 0.105 | 15.868 | 8 | 0.044 |
| | 45-54 | 5,340 | 184.194 | 50 | 0.000 | 1935.949 | 0.034 | 0.103 | 7.543 | 8 | 0.479 |

**Table E.2.** Newly discovered lifestyle risk factors associated with depression ($N$=59140) when demographics are controlled for. All listed lifestyle features meet the significance threshold $p<0.05$ when corrected for multiple comparisons by the Benjamini-Hochberg procedure. Notably, for nominal features, category information is confidential, and we only report results corresponding to the category with the largest Exp($B$) if $B>0$, or the one with the smallest Exp($B$) if $B<0$.

$B$: standardized regression coefficient

$SE$: standard error

$CI$: confidence interval

$p^{BH}$: Benjamini-Hochberg adjusted p-value

| Lifestyle Feature | Coding | Subsample | | | $B$ | $SE$ | Exp($B$) | 95% $CI$ for Exp($B$) | | $p^{BH}$ |
| --- | --- | --- | --- | --- | --- | --- | --- | --- | --- | --- |
| | | Sex | Age | Size | | | | Lower | Upper | |
| Added & Redeemed Coupons | Ordinal (a few → a lot) | Male | 25-34 | 13840 | -0.110 | 0.027 | 0.896 | 0.850 | 0.944 | 0.00030 |
| Alcohol Preference | Nominal (control category: no preference) | Female | 15-24 | 5660 | 2.153 | 0.598 | 8.612 | 2.666 | 27.818 | 0.00493 |
| Auto Parts & Accessories | Ordinal (a few → a lot) | Male | 25-34 | 13840 | -0.177 | 0.041 | 0.838 | 0.773 | 0.908 | 0.00018 |
| Body Weight | Ordinal (thin → fat) | Female | 45-54 | 3900 | 0.214 | 0.071 | 1.239 | 1.077 | 1.424 | 0.02559 |
| Books, Magazines & Newspapers | Ordinal (a few → a lot) | Female | 15-24 | 5660 | 0.158 | 0.055 | 1.171 | 1.052 | 1.304 | 0.02121 |
| | | | 25-34 | 6500 | 0.159 | 0.062 | 1.172 | 1.038 | 1.323 | 0.04292 |
| | | Male | 15-24 | 7400 | 0.140 | 0.051 | 1.151 | 1.041 | 1.272 | 0.03313 |
| | | | 25-34 | 13840 | 0.118 | 0.047 | 1.125 | 1.026 | 1.235 | 0.04897 |
| Camera Phone Preference | Nominal (control category: no preference) | Male | 45-54 | 5340 | -0.465 | 0.158 | 0.628 | 0.461 | 0.856 | 0.02104 |
| Children's Clothing | Ordinal (a few → a lot) | Female | 25-34 | 6500 | -0.155 | 0.060 | 0.856 | 0.762 | 0.963 | 0.04292 |
| | | | 35-44 | 6200 | -0.305 | 0.080 | 0.737 | 0.629 | 0.863 | 0.00173 |

| Lifestyle Feature | Coding | Subsample | | | B | SE | Exp(B) | 95% CI for Exp(B) | | $p^{BH}$ |
|---|---|---|---|---|---|---|---|---|---|---|
| | | Sex | Age | Size | | | | Lower | Upper | |
| | | Male | 25-34 | 13840 | -0.351 | 0.090 | 0.704 | 0.590 | 0.840 | 0.00055 |
| | | | 35-44 | 10300 | -0.254 | 0.095 | 0.776 | 0.644 | 0.935 | 0.04126 |
| Credit Score | Ordinal (bad → good) | Male | 15-24 | 7400 | -0.164 | 0.053 | 0.849 | 0.765 | 0.942 | 0.01530 |
| | | | 25-34 | 13840 | -0.120 | 0.035 | 0.887 | 0.828 | 0.950 | 0.00330 |
| E-Books & Stationery | Ordinal (a few → a lot) | Female | 15-24 | 5660 | 0.140 | 0.047 | 1.150 | 1.048 | 1262. | 0.01838 |
| Financial Status | Ordinal (bad → good) | Female | 15-24 | 5660 | -0.507 | 0.074 | 0.602 | 0.521 | 0.696 | 0.00000 |
| | | Male | 35-44 | 10300 | -0.228 | 0.079 | 0.796 | 0.682 | 0.930 | 0.02678 |
| Haircare Services | Ordinal (a few → a lot) | Female | 15-24 | 5660 | 0.237 | 0.089 | 1.268 | 1.066 | 1.508 | 0.03199 |
| Home Decoration Preference | Nominal (control category: no preference) | Male | 25-34 | 13840 | -0.432 | 0.163 | 0.649 | 0.471 | 0.894 | 0.04843 |
| | | | 35-44 | 10300 | -0.564 | 0.181 | 0.569 | 0.399 | 0.812 | 0.02678 |
| Home Healthcare Supplies | Ordinal (a few → a lot) | Female | 15-24 | 5660 | 0.451 | 0.079 | 1.570 | 1.344 | 1.834 | 0.00000 |
| | | | 25-34 | 6500 | 0.276 | 0.066 | 1.317 | 1.157 | 1.500 | 0.00046 |
| | | | 35-44 | 6200 | 0.349 | 0.070 | 1.418 | 1.235 | 1.628 | 0.00003 |
| | | | 45-54 | 3900 | 0.528 | 0.128 | 1.696 | 1.319 | 2.180 | 0.00147 |
| | | Male | 15-24 | 7400 | 0.385 | 0.063 | 1.469 | 1.298 | 1.663 | 0.00000 |
| | | | 25-34 | 13840 | 0.343 | 0.038 | 1.409 | 1.308 | 1.517 | 0.00000 |
| | | | 35-44 | 10300 | 0.419 | 0.048 | 1.520 | 1.383 | 1.671 | 0.00000 |
| | | | 45-54 | 5340 | 0.453 | 0.063 | 1.573 | 1.389 | 1.780 | 0.00000 |
| Membership Level | Ordinal (low → high) | Male | 15-24 | 7400 | 0.175 | 0.067 | 1.192 | 1.045 | 1.359 | 0.04154 |
| Men's Clothing | Ordinal (a few → a lot) | Female | 25-34 | 6500 | -0.384 | 0.141 | 0.681 | 0.517 | 0.897 | 0.03802 |

| Lifestyle Feature | Coding | Subsample | | | $B$ | $SE$ | $\mathrm{Exp}(B)$ | 95% $CI$ for $\mathrm{Exp}(B)$ | | $p^{\mathrm{BH}}$ |
|---|---|---|---|---|---|---|---|---|---|---|
| | | Sex | Age | Size | | | | Lower | Upper | |
| | | Male | 15-24 | 7400 | -0.298 | 0.079 | 0.743 | 0.636 | 0.867 | 0.00263 |
| Men's Clothing Preference | Nominal (control category: no preference) | Female | 25-34 | 6500 | -0.418 | 0.140 | 0.659 | 0.501 | 0.866 | 0.04292 |
| | | Male | 15-24 | 7400 | -0.467 | 0.127 | 0.627 | 0.489 | 0.804 | 0.00701 |
| Mid-Range Phone Preference | Nominal (control category: no preference) | Female | 15-24 | 5660 | 0.796 | 0.237 | 2.217 | 1.393 | 3.528 | 0.02375 |
| Phone Expenses | Ordinal (a few → a lot) | Female | 15-24 | 5660 | -0.259 | 0.058 | 0.772 | 0.690 | 0.864 | 0.00011 |
| Phone Purchase Price | Ordinal (low → high) | Female | 25-34 | 6500 | -0.147 | 0.040 | 0.863 | 0.798 | 0.934 | 0.00257 |
| | | Male | 15-24 | 7400 | -0.116 | 0.039 | 0.890 | 0.825 | 0.961 | 0.01804 |
| | | | 25-34 | 13840 | -0.109 | 0.027 | 0.896 | 0.851 | 0.945 | 0.00031 |
| Posted Positive Reviews | Ordinal (a few → a lot) | Female | 25-34 | 6500 | -0.386 | 0.091 | 0.680 | 0.568 | 0.813 | 0.00046 |
| | | Male | 15-24 | 7400 | -0.260 | 0.077 | 0.771 | 0.662 | 0.897 | 0.00715 |
| Purchase Amount | Ordinal (low → high) | Female | 15-24 | 5660 | 0.153 | 0.048 | 1.165 | 1.061 | 1.280 | 0.01068 |
| | | | 25-34 | 6500 | 0.178 | 0.051 | 1.195 | 1.081 | 1.321 | 0.00401 |
| | | | 35-44 | 6200 | 0.161 | 0.042 | 1.175 | 1.083 | 1.275 | 0.00163 |
| | | | 45-54 | 3900 | 0.252 | 0.083 | 1.287 | 1.095 | 1.513 | 0.02559 |
| | | Male | 35-44 | 10300 | 0.154 | 0.045 | 1.166 | 1.069 | 1.273 | 0.00778 |
| Purchase Frequency | Ordinal (low → high) | Female | 15-24 | 5660 | 0.234 | 0.067 | 1.263 | 1.108 | 1.440 | 0.00493 |
| | | | 24-34 | 6500 | 0.188 | 0.044 | 1.207 | 1.107 | 1.316 | 0.00046 |
| | | Male | 15-24 | 7400 | 0.159 | 0.030 | 1.173 | 1.106 | 1.244 | 0.00000 |
| | | | 25-34 | 13840 | 0.212 | 0.028 | 1.236 | 1.170 | 1.306 | 0.00000 |
| | | | 35-44 | 10300 | 0.113 | 0.043 | 1.120 | 1.028 | 1.219 | 0.04194 |

| Lifestyle Feature | Coding | Subsample | | | B | SE | Exp(B) | 95% CI for Exp(B) | | $p^{\text{BH}}$ |
|---|---|---|---|---|---|---|---|---|---|---|
| | | Sex | Age | Size | | | | Lower | Upper | |
| | | | 45-54 | 5340 | 0.262 | 0.079 | 1.300 | 1.112 | 1.518 | 0.00830 |
| Second-Hand Goods | Ordinal (a few → a lot) | Male | 15-24 | 7400 | 0.152 | 0.056 | 1.165 | 1.044 | 1.299 | 0.03313 |
| Women's Clothing | Ordinal (a few → a lot) | Female | 25-34 | 6500 | -0.116 | 0.038 | 0.890 | 0.827 | 0.958 | 0.01410 |
| Women's Clothing Preference | Nominal (control category: no preference) | Female | 15-24 | 5660 | -0.826 | 0.247 | 0.438 | 0.270 | 0.711 | 0.01365 |
| | | | 35-44 | 6200 | -0.740 | 0.170 | 0.477 | 0.342 | 0.666 | 0.00162 |
| | | | 45-54 | 3900 | -0.759 | 0.205 | 0.468 | 0.313 | 0.700 | 0.01984 |
| | | Male | 45-54 | 3900 | -0.711 | 0.181 | 0.491 | 0.345 | 0.701 | 0.00578 |

**Table E.3.** Multiple logistic regression (using the enter method) for determining associations between online shoppers' past lifestyle choices and whether they suffer from type 2 diabetes ($N$=38420) when controlling for demographics.

| Subsample | | | Omnibus Tests of Model Coefficients | | | Model Summary | | | Hosmer and Lemeshow Test | | |
|---|---|---|---|---|---|---|---|---|---|---|---|
| Sex | Age | Size | $\chi^2$ | df | p | -2 Log Likelihood | Cox & Snell $R^2$ | Nagelkerke $R^2$ | $\chi^2$ | df | p |
| Female | 15-24 | 2,960 | 340.488 | 112 | 0.000 | 1584.003 | 0.109 | 0.227 | 8.781 | 8 | 0.361 |
|  | 25-34 | 6,400 | 641.162 | 178 | 0.000 | 3519.900 | 0.095 | 0.199 | 8.142 | 8 | 0.420 |
|  | 35-44 | 4,940 | 448.530 | 104 | 0.000 | 2763.289 | 0.082 | 0.182 | 11.011 | 8 | 0.201 |
|  | 45-54 | 2,870 | 193.713 | 48 | 0.000 | 1672.264 | 0.065 | 0.137 | 3.947 | 8 | 0.862 |
|  | 55-64 | 1,160 | 103.897 | 29 | 0.000 | 650.296 | 0.086 | 0.179 | 6.383 | 8 | 0.604 |
| Male | 15-24 | 1,600 | 249.567 | 100 | 0.000 | 790.699 | 0.144 | 0.302 | 15.356 | 8 | 0.053 |
|  | 25-34 | 5,520 | 397.127 | 128 | 0.000 | 3191.789 | 0.069 | 0.145 | 12.041 | 8 | 0.149 |
|  | 35-44 | 5,930 | 449.441 | 88 | 0.000 | 3406.043 | 0.073 | 0.153 | 4.013 | 8 | 0.856 |
|  | 45-54 | 5,040 | 468.274 | 83 | 0.000 | 2808.562 | 0.089 | 0.186 | 12.978 | 8 | 0.113 |
|  | 55-64 | 2,000 | 184.286 | 40 | 0.000 | 1116.046 | 0.088 | 0.184 | 14.173 | 8 | 0.077 |

**Table E.4.** Newly discovered lifestyle risk factors associated with type 2 diabetes (*N*=38420) when demographics are controlled for. All listed lifestyle features meet the significance threshold *p*<0.05 when corrected for multiple comparisons by the Benjamini-Hochberg procedure. Notably, for nominal features, category information is confidential, and we only report results corresponding to the category with the largest Exp(*B*) if *B*>0, or the one with the smallest Exp(*B*) if *B*<0.

*B*: standardized regression coefficient

*SE*: standard error

*CI*: confidence interval

*p*[BH]: Benjamini-Hochberg adjusted p-value

| Lifestyle Feature | Coding | Subsample | | | *B* | *SE* | Exp(*B*) | 95% *CI* for Exp(*B*) | | *p*[BH] |
|---|---|---|---|---|---|---|---|---|---|---|
| | | Sex | Age | Size | | | | Lower | Upper | |
| Body Weight | Ordinal (thin → fat) | Female | 25-34 | 6400 | 0.105 | 0.023 | 1.111 | 1.063 | 1.161 | 0.00012 |
| | | | 35-44 | 4940 | 0.096 | 0.032 | 1.101 | 1.034 | 1.173 | 0.03054 |
| Bodycare Products | Ordinal (a few → a lot) | Female | 35-44 | 4940 | 0.235 | 0.065 | 1.264 | 1.113 | 1.437 | 0.00601 |
| Cereals, Dried Foods & Condiments | Ordinal (a few → a lot) | Female | 15-24 | 2960 | 0.187 | 0.065 | 1.205 | 1.062 | 1.369 | 0.04674 |
| | | | 25-34 | 6400 | 0.130 | 0.045 | 1.139 | 1.044 | 1.243 | 0.03013 |
| | | Male | 25-34 | 5520 | 0.190 | 0.059 | 1.210 | 1.077 | 1.359 | 0.01294 |
| Children's Clothing | Ordinal (a few → a lot) | Male | 35-44 | 5930 | -0.269 | 0.098 | 0.764 | 0.630 | 0.927 | 0.02870 |
| Coffee, Oatmeals & Powdered Drink Mixes | Ordinal (a few → a lot) | Male | 55-64 | 2000 | 0.320 | 0.106 | 1.377 | 1.120 | 1.694 | 0.01226 |
| Coffee Preference | Nominal (control category: no preference) | Female | 15-24 | 2960 | 0.786 | 0.303 | 2.195 | 1.211 | 3.979 | 0.04163 |
| Credit Score | Ordinal (bad → good) | Male | 25-34 | 5520 | -0.127 | 0.045 | 0.881 | 0.806 | 0.962 | 0.03891 |
| | | | 35-44 | 5930 | -0.125 | 0.044 | 0.883 | 0.810 | 0.962 | 0.02578 |

| Lifestyle Feature | Coding | Subsample | | | B | SE | Exp(B) | 95% CI for Exp(B) | | $p^{BH}$ |
|---|---|---|---|---|---|---|---|---|---|---|
| | | Sex | Age | Size | | | | Lower | Upper | |
| E-Books & Stationery | Ordinal (a few → a lot) | Male | 45-54 | 5040 | -0.234 | 0.093 | 0.792 | 0.660 | 0.950 | 0.04177 |
| Energy-Hungry Appliance Preference | Nominal (control category: no preference) | Female | 25-34 | 6400 | 2.062 | 0.663 | 7.862 | 2.143 | 28.841 | 0.00317 |
| Financial Status | Ordinal (bad → good) | Female | 15-24 | 2960 | -0.438 | 0.106 | 0.617 | 0.501 | 0.760 | 0.00012 |
| | | Female | 25-34 | 6400 | -0.340 | 0.079 | 0.712 | 0.610 | 0.830 | 0.00035 |
| | | Male | 15-24 | 1600 | -0.609 | 0.149 | 0.544 | 0.406 | 0.728 | 0.00071 |
| Home Decoration Preference | Nominal (control category: no preference) | Female | 25-34 | 6400 | -0.828 | 0.362 | 0.437 | 0.215 | 0.888 | 0.01271 |
| | | Male | 25-34 | 5520 | -0.551 | 0.160 | 0.576 | 0.421 | 0.788 | 0.00102 |
| | | Male | 35-44 | 5930 | -0.612 | 0.176 | 0.542 | 0.384 | 0.766 | 0.00168 |
| | | Male | 45-54 | 5040 | -0.604 | 0.220 | 0.547 | 0.355 | 0.842 | 0.00373 |
| Home Healthcare Supplies | Ordinal (a few → a lot) | Female | 15-24 | 2960 | 0.970 | 0.105 | 2.639 | 2.148 | 3.242 | 0.00000 |
| | | Female | 25-34 | 6400 | 0.487 | 0.058 | 1.627 | 1.453 | 1.822 | 0.00000 |
| | | Female | 35-44 | 4940 | 0.687 | 0.069 | 1.988 | 1.737 | 2.275 | 0.00000 |
| | | Female | 45-54 | 2870 | 0.816 | 0.105 | 2.261 | 1.842 | 2.776 | 0.00000 |
| | | Female | 55-64 | 1160 | 0.668 | 0.132 | 1.950 | 1.505 | 2.526 | 0.00001 |
| | | Male | 15-24 | 1600 | 0.848 | 0.171 | 2.335 | 1.670 | 3.264 | 0.00002 |
| | | Male | 25-34 | 5520 | 0.510 | 0.077 | 1.665 | 1.432 | 1.935 | 0.00000 |
| | | Male | 35-44 | 5930 | 0.740 | 0.069 | 2.095 | 1.830 | 2.399 | 0.00000 |
| | | Male | 45-54 | 5040 | 0.925 | 0.094 | 2.522 | 2.097 | 3.032 | 0.00000 |
| | | Male | 55-64 | 2000 | 0.574 | 0.086 | 1.776 | 1.499 | 2.103 | 0.00000 |
| Home Improvements | Ordinal (a few → a lot) | Male | 45-54 | 5040 | -0.756 | 0.189 | 0.469 | 0.324 | 0.679 | 0.00063 |

| Lifestyle Feature | Coding | Subsample | | | $B$ | $SE$ | $Exp(B)$ | 95% $CI$ for $Exp(B)$ | | $p^{BH}$ |
|---|---|---|---|---|---|---|---|---|---|---|
| | | Sex | Age | Size | | | | Lower | Upper | |
| Kitchen Appliances | Ordinal (a few → a lot) | Female | 45-54 | 2870 | 0.395 | 0.127 | 1.484 | 1.158 | 1.902 | 0.03103 |
| Large-Screen Phone Preference | Nominal (control category: no preference) | Male | 45-54 | 5040 | 0.484 | 0.186 | 1.622 | 1.126 | 2.335 | 0.03999 |
| Meats, Seafoods & Vegetables | Ordinal (a few → a lot) | Female | 25-34 | 6400 | 0.113 | 0.040 | 1.120 | 1.035 | 1.212 | 0.03510 |
| Membership Level | Ordinal (low → high) | Female | 25-34 | 6400 | 0.112 | 0.042 | 1.118 | 1.030 | 1.214 | 0.04306 |
| | | Male | 25-34 | 5520 | 0.142 | 0.044 | 1.152 | 1.058 | 1.256 | 0.01294 |
| | | | 35-44 | 5930 | 0.114 | 0.041 | 1.120 | 1.033 | 1.215 | 0.02870 |
| Mid-Range Phone Preference | Nominal (control category: no preference) | Male | 45-54 | 5040 | 0.590 | 0.185 | 1.805 | 1.255 | 2.594 | 0.04935 |
| Phone Purchase Price | Ordinal (low → high) | Female | 25-34 | 6400 | -0.086 | 0.031 | 0.917 | 0.863 | 0.975 | 0.03510 |
| | | | 35-44 | 4940 | -0.134 | 0.036 | 0.874 | 0.815 | 0.938 | 0.00499 |
| Posted Positive Reviews | Ordinal (a few → a lot) | Male | 45-54 | 5040 | -0.357 | 0.114 | 0.700 | 0.560 | 0.874 | 0.00751 |
| Purchase Amount | Ordinal (low → high) | Female | 35-44 | 4940 | 0.129 | 0.046 | 1.137 | 1.040 | 1.244 | 0.04637 |
| Purchase Frequency | Ordinal (low → high) | Male | 35-44 | 5930 | 0.198 | 0.037 | 1.219 | 1.134 | 1.311 | 0.00000 |
| | | | 45-54 | 5040 | 0.253 | 0.041 | 1.288 | 1.188 | 1.395 | 0.00000 |
| | | | 55-64 | 2000 | 0.264 | 0.058 | 1.302 | 1.161 | 1.460 | 0.00006 |
| Purchased Product Categories | Ordinal (a few → a lot) | Male | 45-54 | 5040 | -0.334 | 0.105 | 0.716 | 0.583 | 0.880 | 0.00751 |
| Second-Hand Goods | Ordinal (a few → a lot) | Male | 15-24 | 1600 | 0.542 | 0.187 | 1.719 | 1.191 | 2.482 | 0.04333 |
| Spending Per Purchase | Ordinal (low → high) | Female | 25-34 | 6400 | 0.171 | 0.056 | 1.186 | 1.064 | 1.323 | 0.02162 |
| | | Male | 25-34 | 5520 | 0.182 | 0.051 | 1.200 | 1.086 | 1.326 | 0.00561 |
| Tableware | Ordinal (a few → a lot) | Female | 25-34 | 6400 | -0.198 | 0.071 | 0.820 | 0.713 | 0.944 | 0.03510 |
| Tea Beverages | Ordinal (a few → a lot) | Male | 45-54 | 5040 | 0.169 | 0.050 | 1.184 | 1.073 | 1.306 | 0.00481 |

| Lifestyle Feature | Coding | Subsample | | | $B$ | $SE$ | $\text{Exp}(B)$ | 95% $CI$ for $\text{Exp}(B)$ | | $p^{\text{BH}}$ |
|---|---|---|---|---|---|---|---|---|---|---|
| | | Sex | Age | Size | | | | Lower | Upper | |
| Tier Of City Of Residence | Ordinal (urban → rural) | Female | 25-34 | 6400 | -0.133 | 0.043 | 0.875 | 0.804 | 0.953 | 0.02162 |
| | | | 35-44 | 4940 | -0.173 | 0.051 | 0.841 | 0.762 | 0.929 | 0.00894 |
| Women's Clothing | Ordinal (a few → a lot) | Male | 35-44 | 5930 | -0.374 | 0.107 | 0.688 | 0.558 | 0.848 | 0.00404 |
| Women's Clothing Preference | Nominal (control category: no preference) | Female | 35-44 | 4940 | -0.438 | 0.144 | 0.645 | 0.486 | 0.856 | 0.04654 |
| | | | 55-64 | 1160 | -0.991 | 0.246 | 0.371 | 0.229 | 0.602 | 0.00053 |
| | | Male | 35-44 | 5930 | -0.411 | 0.122 | 0.663 | 0.522 | 0.842 | 0.01115 |
| | | | 55-64 | 2000 | -0.898 | 0.257 | 0.407 | 0.246 | 0.674 | 0.01226 |

# Appendix F. Performance Evaluation

**Table F.1.** Online shopping behaviors-based prediction performance (via SVM) in early risks of depression ($N$=8) and type 2 diabetes ($N$=10) when demographics are controlled for. The evaluation metrics include sensitivity, specificity, positive predictive value (PPV), negative predictive value (NPV), accuracy, $F_1$ score, and area under the receiver operating characteristic curve (AUC), reported as cross-validated out-of-sample mean ± standard deviation.

| Evaluation metric | Depression | Type 2 Diabetes |
|:---:|:---:|:---:|
| Sensitivity | 0.611±0.037 | 0.553±0.069 |
| Specificity | 0.639±0.043 | 0.741±0.068 |
| PPV | 0.083±0.008 | 0.200±0.030 |
| NPV | 0.969±0.002 | 0.937±0.005 |
| Accuracy | 0.638±0.040 | 0.722±0.055 |
| $F_1$ score | 0.145±0.012 | 0.288±0.023 |
| AUC | 0.678±0.025 | 0.695±0.025 |

# Appendix G. Comparison Methods

**Table G.1.** An electronic medical records-based detection method of depression proposed by Trinh et al. [1]

| EMR Field | Predictors | Prediction Performance* | | | | |
|---|---|---|---|---|---|---|
| | | Sensitivity | Specificity | AUC | PPV | NPV |
| Diagnostic Code | ICD-9 Codes, including 290.13, 290.21, 290.43, 296.2, 296.3, 296.82, 296.9, 296.99, 298, 300.4, 301.1, 305.8, 305.81, 309, 309.1, 311, 969, and E939.0; 11 Hospital-Specific Billing Codes | 0.77 | 0.76 | 0.77 | 0.76 | 0.77 |
| Problem List | Any Mention of "Depression", "Depressed", and "Major Depression" | 0.49 | 0.78 | 0.63 | 0.68 | 0.61 |
| Medication List | Antidepressants, including Amitriptyline, Fluoxetine, Phenelzine, Bupropion, Fluvoxamine, Protriptyline, Citalopram, Imipramine, Sertraline, Clomipramine, Maprotiline, Selegiline Patch, Desipramine, Mirtazapine, Tranylcypromine, Doxepin, Nefazodone, Trimipramine, Duloxetine, Nortriptyline, Venlafaxine, Escitalopram, and Paroxetine | 0.56 | 0.88 | 0.72 | 0.83 | 0.67 |
| Combination of All EMR Fields | | 0.25 | 0.96 | 0.61 | 0.86 | 0.57 |

\* tested on 427 primary care patients from the Massachusetts General Hospital, the United States

EMR: electronic medical record

AUC: area under the receiver operating characteristic curve

PPV: positive predictive value

NPV: negative predictive value

BMI: body mass index

**Table G.2.** Seven diabetes risk assessment instruments tested by Gao et al. [2]

| Baseline | Predictors | Prediction Performance* | | |
|---|---|---|---|---|
| | | Sensitivity | Specificity | AUC |
| Cambridge Risk Model [3] | Age, Sex, Drug-Treated Hypertension, Corticosteroid Treatment, Family History of Diabetes, BMI, and Smoking | 0.422 | 0.795 | 0.676 |
| Danish Risk Score [4] | Age, Sex, BMI, Known History of Hypertension, Diabetes in Parents, and Physical Activity at Leisure Time | 0.551 | 0.721 | 0.690 |
| Indian Risk Score [5] | Age, Sex, Family History of Diabetes, BMI, Waist Circumference, and Physical Activity | 0.961 | 0.187 | 0.675 |
| Rotterdam Study [6] | Age, Sex, Drug-Treated Hypertension, and BMI | 0.188 | 0.904 | 0.631 |
| Finnish Risk Score [7] | Age, BMI, Waist, Drug-Treated Hypertension, Physical Inactivity, and Daily Consumption of Vegetables, Fruit or Berries | 0.395 | 0.804 | 0.665 |
| Thai Risk Score [8] | Age, Sex, BMI, Waist Circumference, Hypertension, and Family History of Diabetes | 0.868 | 0.326 | 0.662 |
| Chinese Risk Score [2] | Age, Waist Circumference, and Diabetes in Parents or Siblings | 0.842 | 0.398 | 0.673 |

* tested on 4336 participants of the 2006 diabetes survey of Qingdao, China

AUC: area under the receiver operating characteristic curve

BMI: body mass index

# References

1. Trinh NHT, Youn SJ, Sousa J, et al. Using electronic medical records to determine the diagnosis of clinical depression. *Int J Med Inform* 2011; 80(7): 533-540.
2. Gao WG, Dong YH, Pang ZC, et al. A simple Chinese risk score for undiagnosed diabetes. *Diabet Med* 2010; 27(3): 274-281.
3. Engelgau MM, Narayan KM, and Herman WH. Screening for type 2 diabetes. *Diabetes Care* 2000; 23(10): 1563-1580.
4. Glümer C, Carstensen B, Sandbæk A, et al. A Danish diabetes risk score for targeted screening: the Inter99 study. *Diabetes Care* 2004; 27(3): 727-733.
5. Ramachandran A, Snehalatha C, Vijay V, et al. Derivation and validation of diabetes risk score for urban Asian Indians. *Diabetes Res Clin Pract* 2005; 70(1): 63-70.
6. Baan CA, Ruige JB, Stolk RP, et al. Performance of a predictive model to identify undiagnosed diabetes in a health care setting. *Diabetes Care* 1999; 22(2): 213-219.
7. Lindström J and Tuomilehto J. The diabetes risk score: a practical tool to predict type 2 diabetes risk. *Diabetes Care* 2003; 26(3): 725-731.
8. Aekplakorn W, Bunnag P, Woodward M, et al. A risk score for predicting incident diabetes in the Thai population. *Diabetes Care* 2006; 29(8): 1872-1877.